\documentstyle[preprint,aps]{revtex}
\begin{document}

\draft


\title{Quantisation of Conformal Fields in Three-dimensional \\ 
 Anti-de Sitter Black Hole Spacetime}

\author{Hongsu Kim$^{a}$, Jae-Sok Oh$^{b}$ and Changrim Ahn$^{c}$}

\address{$a,~c$ Department of Physics\\
Ewha Women's University, Seoul 120-750, KOREA \\
$b$ Department of Physics\\
Sogang University, C.P.O. Box 1142, Seoul 100-611, KOREA}

\date{August, 1998}

\maketitle

\begin{abstract}
Utilizing the conformal-flatness nature of 3-dim. Anti-de Sitter
(AdS$_3$) black hole solution of Banados, Teitelboim and Zanelli,
the quantisation of conformally-coupled scalar and spinor fields
in this background spacetime is explicitly carried out. In particular,
mode expansion forms and propagators of the fields are obtained in
closed forms. The vacuum in this conformally-coupled field theories  
in AdS$_3$ black hole spacetime, which is conformally-flat, is the 
conformal vacuum which is unique and has global meaning. 
This point particularly suggests that now the particle production by
AdS$_3$ black hole spacetime should be absent. General argument 
establishing the absence of real particle creation by AdS$_3$ black 
hole spacetime for this case of conformal triviality is provided.
Then next, using the explicit mode expansion forms for 
conformally-coupled scalar and spinor fields, the bosonic and fermionic
superradiances are examined and found to be absent confirming the 
expectation. 
\end{abstract}


\narrowtext


\newpage

\noindent
\begin{center}
{\rm\bf I. Introduction}
\end{center}

We begin with some comments on peculiar features of gravity in 
(2+1)-dimensions. They are the {\it local flatness}
and the {\it absence of conformal anomaly}. Firstly on the local 
flatness ; consider the Einstein field equation in an arbitrary 
$n( n \ge 3)$ dimensions (our sign convention here is chosen to be that of 
Misner,Thorne and Wheeler [1])
\begin{eqnarray}
  G_{\mu\nu} = R_{\mu\nu} - {1\over2}g_{\mu\nu}R = 8{\pi}GT_{\mu\nu}.
\end{eqnarray}
Now, generally the vanishing of $ R_{\mu\nu}$ and $ R$ and hence of the 
Einstein tensor $ G_{\mu\nu}$ (due to the absence of the matter, 
$T_{\mu\nu}=0 $) does not necessarily imply the vanishing Riemann tensor,
$R_{\mu\nu\alpha\beta}$ although the converse is true. Namely, the empty 
space needs not be flat. On the contrary in 3-dimensions, the relation 
$G^{\mu}_{\nu} = -{1\over4}\epsilon^{\mu\alpha\beta}
\epsilon_{\nu\gamma\delta}R_{\alpha\beta}^{\gamma\delta}$, when 
inverted, yields   $ R^{\alpha\mu}_{\beta\nu} 
= -\epsilon^{\alpha\mu\rho}\epsilon_{\beta\nu\sigma}G^{\sigma}_{\rho}$
implying that the Riemann tensor is directly proportional to the Einstein 
tensor or the matter energy-momentum tensor.  As a result, the vanishing 
of $G_{\mu\nu}$ due to the absense of matter necessarily implies the vanishing 
of $R_{\mu\nu\alpha\beta}$.  This indicates that 3-dimensional empty 
space is necessarily flat. Several consequenses follow immediately.  Since 
the vacuum spacetime is locally-flat, there are no gravitational waves in 
the classical theory and upon quantization, there are no quantum gravitons.  
Matter sources may produce curvature but only locally at the location of 
the sources.  And the forces, if any, between sources are not mediated by 
graviton exchange since there are no gravitons.  This also means that the 
Newtonian limit of general relativity is lost in 3-dimensions and we are left 
with an apparently uninteresting theory.  In spite of these frustrating 
observations, the gravity in 3-dimensions is not completely devoid of physical 
relevances.  The vanishing of Riemann tensor means that any point in a 
spacetime manifold M has a neighborhood that is isometric to the Minkowski 
spacetime.  Thus if M has a trivial topology, a single neighborhood can be 
extended globally and the geometry is indeed trivial (flat).  But if M has 
a non-trivial topology, say, it contains noncontractable curves, such an 
extension may not be possible and as a consequence interesting physics may 
emerge.  To name a few, such existence of  ``global geometry" for spacetimes 
with nontrivial topologies may arise by coupling point particles 
(both with and without spin) to gravity or by adding a cosmological 
constant to the Einstein field equation [3]. Particularly, the elaboration 
of adding a cosmological constant to the vacuum Einstein theory generated
much excitement recently. Namely, due to this peculiar aspect of gravity
in 3-dimensions, it had long been thought that black hole solutions cannot
exist in 3-dimensions since there is no local gravitational attraction
and hence no mechanism to confine large densities of matter. It was, therefore,
quite a surprise when Banados, Teitelboim and Zanelli (BTZ) have recently
constructed the Anti-de Sitter (AdS$_3$) spacetime solutions to 3-dimensional
Einstein equation that can be interpreted as black hole solutions. They
included the negative cosmological constant in the 3-dimensional vacuum
Einstein theory and then found both rotating and nonrotating black hole 
solutions. \\ 
Secondly, we turn to the general features of ``conformal flatness" 
of spacetimes which 
turns out to be a great advantage that allows us to carry out calculations, 
in full detail, involved in dealing with the conformally-coupled
quantum fields propagating in the  background of
conformally-flat spacetimes. 
Note that generally the Riemann tensor $R_{\mu\nu\alpha\beta}$, 
containing the full information about the curvature of  spacetime, may be 
expressed, for dimensions $ n\ge 3$, in terms of its various traces given by 
the Ricci tensor $R_{\mu\nu}$, the curvature scalar R and the traceless, 
conformally-invariant piece, the Weyl or conformal tensor,  
$C_{\mu\nu\alpha\beta}$ as [10]

\begin{eqnarray}
 R_{\mu\nu\alpha\beta} 
    &=& {1\over(n-2)}(  g_{\mu\alpha}R_{\nu\beta} 
                      + g_{\nu\beta} R_{\mu\alpha} 
                      - g_{\mu\beta} R_{\nu\alpha} 
                      - g_{\nu\alpha}R_{\mu\beta})       \nonumber      \\
    &-& {R\over{(n-1)(n-2)}}(g_{\mu\alpha}g_{\nu\beta}   
                           - g_{\mu\beta}g_{\nu\alpha}) 
                           + C_{\mu\nu\alpha\beta}.
\end{eqnarray}
The Weyl tensor, as is well known, has a dual role;  not only is it the 
traceless part of the Riemann tensor, it also ``probes" the conformal 
properties of a metric.  First, $C_{\mu\nu\alpha\beta}$ is invariant 
under conformal transformation of the metric, 
$g_{\mu\nu} = \Omega^2(x)\tilde{g}_{\mu\nu}$. Next, but more
importantly, it vanishes if and only if the metric is conformally-flat,  
$g_{\mu\nu} = \Omega^2(x)\eta_{\mu\nu}$ for spacetimes with dimensions
$n>3$.  In $\it n$=3-dimensions, however, 
the Weyl tensor vanishes ``identically''  $C_{\mu\nu\alpha\beta} = 0$. 
In particular, the very consequence of the vanishing Weyl tensor in 
$n=3$, i.e.,   

\begin{eqnarray}
 R_{\mu\nu\alpha\beta} 
    &=& (g_{\mu\alpha}R_{\nu\beta} 
       + g_{\nu\beta} R_{\mu\alpha} 
       - g_{\mu\beta} R_{\nu\alpha}
       - g_{\nu\alpha}R_{\mu\beta})                    \nonumber \\
    && -{1\over2}R(g_{\mu\alpha}g_{\nu\beta}  
                 - g_{\mu\beta} g_{\nu\alpha}) 
\end{eqnarray}
establishes the fact that (along with the fact that both the Riemann 
tensor and Ricci tensor have the same number (6) of independent 
components) the Riemann tensor is linearly proportional to the Ricci and 
hence to the Einstein tensor,  $R^{\alpha\mu}_{\beta\nu} 
= -\epsilon^{\alpha\mu\rho}\epsilon_{\beta\nu\sigma}G^{\sigma}_{\rho}$ as 
stated earlier.  Since the Weyl tensor $C_{\mu\nu\alpha\beta}$ vanishes
identically in $n=3$, one now needs some other means to probe the
conformal flatness of 3-dim. spacetimes. Indeed it is known that in
$n=3$, the ``Weyl-Schouten'' tensor [11] defined by
\begin{eqnarray}
C_{\lambda\mu\nu} \equiv \nabla_{\nu}R_{\lambda\mu} - \nabla_{\mu}
R_{\lambda\nu} - {1\over 4}(g_{\lambda\mu}\partial_{\nu}R - 
g_{\lambda\nu}\partial_{\mu}R) \nonumber
\end{eqnarray}
plays the role of Weyl tensor, namely $C_{\lambda\mu\nu}=0$
if and only if a 3-dim. spacetime is conformally flat.
The AdS$_3$ black hole spacetime we mentioned earlier is a solution
to the Einstein equation
$R_{\mu\nu}-{1\over 2}g_{\mu\nu}R+\Lambda g_{\mu\nu} = 0$ which, in
turn, implies $R_{\mu\nu}=2\Lambda g_{\mu\nu}$ and hence
$R=6\Lambda$. Namely, since its scalar curvature is constant and its
Ricci tensor is covariantly constant, clearly $C_{\lambda\mu\nu}=0$
and thus this AdS$_3$ black hole spacetime is conformally-flat.
As a matter of fact, this conclusion was expected since we know that
generally, (anti) de Sitter spacetimes are spaces of constant curvature
and hence are conformally-flat. In this work, we 
shall first present a general formulation for the quantisation of 
``conformally-coupled" scalar and spinor fields (but not the vector field 
which fails to remain conformally-coupled in dimensions other than four) 
in the background of the conformally-flat $n$-dimensional spacetimes.  
Then, as a particular and interesting example, we shall take the 
AdS$_3$ black hole spacetime of BTZ, which is also conformally flat, 
and perform 
a quantisation of conformally-coupled matter fields on this background. \\
Lastly, on the absence of conformal anomaly ; it is natural to wonder
if the conformal anomaly necessarily comes into play when considering the
quantisation of conformally-coupled scalar and spinor fields on 
conformally-flat AdS$_3$ black hole spacetime
which is of our particular interest. The answer to this question is obviously 
``no'' and one needs not worry about the conformal anomaly in the first place.
The reason for this goes as follows ; as is 
well-known, the conformal anomaly is closely related to the ``trace
anomaly'' [2]. Namely the non-vanishing trace of the renormalized stress
tensor leads to the conformal non-invariance of the quantum effective
action of the matter field since
\begin{eqnarray}
< T^{\mu}_{ren ~\mu} (x) > = {2\over \sqrt{g}} g_{\mu\nu} 
{\delta W_{ren} \over \delta g_{\mu\nu}}.  \nonumber
\end{eqnarray}
Now in the context of zeta function regularization scheme, for example,
the renormalized 1-loop effective action is given by [2,15]
$W_{ren} = -{1\over 2}\xi' (0)$ for real scalar field and
$W_{ren} = \xi' (0)$ for spinor field in Euclidean signature. Then the 
explicit calculation demonstrates that $< T^{\mu}_{ren ~\mu} (x) >$
is non-zero in even dimensions ($d = 2, 4$) while vanishing in 
odd-dimensions (including $d = 3$). Namely the conformal anomaly
emerges only in even dimensions and is absent in odd spacetime 
dimensions. And this general statement holds true for conformally-flat
spacetimes such as AdS$_3$ black hole spacetime as well. \\
Now, speaking of the motivations of the present work, as already
mentioned, partly we wish to present a standard formulation of the general
quantization of conformally-coupled scalar and spinor fields in 
conformally-flat spacetimes. And partly we would like to clarify an 
unsatisfactory state of affair concerning the existing study of 
thermodynamics of 3-dimensional AdS$_3$ black hole of BTZ.
Let us be more specific on this last point.
Essentially, one is bound to be left with the Hawking evaporation and hence
the thermodynamics of black holes when he/she considers minimally-coupled
(massive or massless) or non-minimally coupled (including the conformal
coupling) quantum fields in the background of 4-dimensional black hole
spacetimes. Thus, in principle, one may employ any type of matter coupling
(to background gravity) to treat the thermodynamics of 4-dimensional
black holes although, in practice, general couplings involve highly
non-trivial complications. Perhaps because of this state of matter, people
do not seem to be cautious in selecting the matter coupling to consider
the thermodynamics of black holes in lower dimensions, particularly
those in 3-dimensions such as the AdS$_3$ black holes of BTZ. Unlike
the 4-dimensional case, however,
the AdS$_3$ black hole spacetimes is conformally-flat.
Therefore if one negligently employs conformally-coupled quantum fields
in the AdS$_3$ black hole spacetimes
(which is of our interest here) to discuss its thermodynamics, 
he/she may be misled and run into a trouble. What happens is, in this
case of ``conformal triviality'' (i.e., when fields are conformally-coupled
to the conformally-flat spacetimes), the associated Fock vacuum is the
``conformal vacuum'' which is unique and has global meaning and hence 
there is no real particle creation. As a consequence, the Hawking
temperature (but not the {\it local} temperature measured by an 
accelerating detector) is zero and thus the discussion of black hole
thermodynamics becomes irrelevant to start with. Namely, in order to
carry out a meaningful examination of AdS$_3$ black holes 
thermodynamics, one should employ
non-conformal matter couplings. Indeed, such confusion regarding the
study of thermodynamics of BTZ black holes appeared in the literature [5]
and our present work is partly intended to clarify this issue.
To be more concrete, for example, the authors of ref. 5 employed
conformally-coupled scalar and spinor fields to point out correctly
the ``statistical inversion'' arising in the accelerating detector's
response function but to discuss incorrectly the Hawking temperature
and black hole thermodynamics which become irrelevant because of the
conformally trivial setting. In the present work (particularly in 
sect.IV), we shall provide a general argument and evidences that 
explain this seemingly contradictory phenomenon, i.e., the presence 
of ``particle detection'' but the absence of real `particle creation''
in the case of conformal triviality. \\
This paper is organized as follows :
In sect. II, we formally review the formalism of quantizing conformally-
coupled scalar and spinor fields in general $n$-dimensional conformally 
flat spacetimes.
In sect. III, as an application of the general formalism given in sect. II,
the mode expansions of conformally-coupled quantum fields in the AdS$_3$
black hole spacetime of BTZ are obtained in closed forms. 
Sect. IV will be devoted
to the general argument on the absence of particle creation in conformal
triviality which is the case at hand. In sect. V, as an explicit
demonstration of the absence of real particle creation, the absence of
the superradiant scatterings off the AdS$_3$ black hole is shown using
the explicit mode expansion forms obtained in sect. III.
Finally in sect. VI, we summarize the results of our present work. 
In addition in the appendix, Green's functions of conformally-coupled
quantum fields in the background of AdS$_3$ black hole spacetime are 
given in closed forms as well.


\vspace*{0.5cm}

\noindent
\begin{center}
{\rm\bf II. General Formulation}
\end{center}
{\rm\bf 1. Quantization of conformally-coupled scalar field in 
conformally-flat spacetimes}
\\
Although we are particularly interested in the quantisation of 
conformally-coupled fields in conformally-flat 3-dimensional spacetimes, 
in what follows we shall formulate the quantisation scheme generally in 
conformally-flat $\it n$-dim. spacetimes [2].  
The action for a real scalar field coupled conformally 
to gravity is given by [2] (assuming the gravity as a ``background")

\begin{eqnarray}
   S = -\int d^{n}x\sqrt{g}
        [{1\over2}g^{\mu\nu}\partial_{\mu}\Phi\partial_{\nu}\Phi 
       + {1\over2}\xi(n)R\Phi^2] 
\end{eqnarray}
where $\xi(n) = {(n-2)\over4(n-1)}$ in $\it n$-dimensions. 
One can readily check that this action is invariant under the Weyl-rescaling
 
\begin{eqnarray}
  g_{\mu\nu} = \Omega^2(x)\tilde{g}_{\mu\nu},\qquad 
        \Phi = \Omega^{-({{n-2}\over2})}\tilde{\Phi}         
\end{eqnarray}              
In particular, if the background spacetime is ``conformally flat", 
$g_{\mu\nu} = \Omega^2(x)\eta_{\mu\nu}$, 
then from $\tilde{g}_{\mu\nu} = \eta_{\mu\nu}$, it follows that 
$\tilde{\Box} = \eta^{\mu\nu}\partial_{\mu}\partial_{\nu}$,\ 
$\tilde{R}(\tilde{g}) = 0$ and hence, we are left with the action
\begin{eqnarray}
 S = -\int d^{n}x{1\over2}\tilde{\Phi}(-\tilde{\Box})\tilde{\Phi} 
   = -\int d^{n}x{1\over2}\eta^{\mu\nu}\partial_{\mu}\tilde{\Phi} 
      \partial_{\nu}\tilde{\Phi}.  
\end{eqnarray}
This implies that the theory of conformally-coupled scalar field in 
conformally flat spacetime can be substituted with the usual 
free, massless scalar field theory in flat spacetime of the Weyl-rescaled 
field, $\tilde{\Phi}(x) = \Omega^{({{n-2}\over2})}(x)\Phi(x)$ with 
$\Omega(x)$ being the spacetime dependent conformal factor. 
Thus in order to quantize the original field $\Phi(x)$, we first carry 
out the familiar quantisation programme for the free, Weyl-rescaled field  
$\tilde{\Phi}(x)$ and then use the result to eventually recover the 
quantized theory of the original field  $\Phi(x)$ via  
$\Phi(x) = \Omega^{-({{n-2}\over2})}(x)\tilde{\Phi}(x)$ at the end.  
For instance, if we employ the standard canonical quantisation scheme, 
we begin by demanding the equal time commutators
\begin{eqnarray}
 [\tilde{\Phi}(t,\vec{x}), \tilde{\Pi}(t,\vec{y})] 
  &=& i\delta^{n-1}(\vec{x}  - \vec{y}),        \\
 {[\tilde{\Phi}(t,\vec{x}), \tilde{\Phi}(t,\vec{y})]} 
  &=& [\tilde{\Pi}(t,\vec{x}), \tilde{\Pi}(t,\vec{y})] = 0. \nonumber
\end{eqnarray}
where
$\tilde{\Pi}(x) = {\delta{S}}/{\delta(\partial_t\tilde{\Phi})}
                = \partial_t\tilde{\Phi}$
is the momentum conjugate to the Weyl-rescaled scalar field.
Then for the particle interpretation, we decompose the scalar field using 
the mode expansion

\begin{eqnarray}
 \tilde{\Phi}(x) 
   &=& \int {{d^{n-1}k\over{[(2\pi)^{n-1}2\omega_k]^{1/2}}}}
       [a(k)\tilde{u}_k(x) + a^{\dagger}(k)\tilde{u}^{*}_k(x)]   
       \qquad (\omega_k{\equiv}k^0)                \nonumber  \\
   &{\equiv}&\sum\limits_{\vec{k}}\; 
       [a_{k}\tilde{u}_k(x) + a^{\dagger}_k\tilde{u}^{*}_k(x)], \nonumber
\end{eqnarray}
where the mode functions, which are the solutions of the Klein-Gordon
equation
$\partial_{\mu}\partial^{\mu}\tilde{\Phi} = \Box\tilde{\Phi} = 0$, are
given by 
\begin{eqnarray}
 \tilde{u}_{k}(x) 
    = {{1\over{[(2\pi)^{n-1}2\omega_{k}]^{1\over2}}}}e^{ik{\cdot}x}    
\end{eqnarray}
and $a^{\dagger}(k)$ and $a(k)$ are the creation and the annihilation
operators respectively.
Now, then the mode expansion of the original field is recovered as

\begin{eqnarray}
  \Phi(x) = \Omega^{-({{n-2}\over2})}(x)\sum\limits_{\vec k} 
         [a_{k}\tilde{u}_{k}(x) + a^{\dagger}_{k}\tilde{u}^{*}_{k}(x)].  
\end{eqnarray}
Note, here, that the conformal factor $\Omega(x)$ is not a field which
is subject to the quantisation nor it participates in the physical
particle interpretation.
The vacuum state associated with these mode function of the original 
field above, namely, $a_{k}|0> = 0$ is called the ``conformal vacuum" [2].  
The physical meaning of this conformal vacuum may be described as follows;
as is well-known, generally in curved spacetimes, a particular set of 
mode functions of a field equation and the corresponding vacuum and Fock 
space do not in general have direct physical significance.  Or put in 
plain English, in curved spacetimes, there is in general no meaningful 
notion of global vacuum state.  A vacuum state in a reference frame may 
be a many-particle state in another.  The concept of particle is really 
observer-dependent.  Nevertheless, if there exist geometrical symmetries 
in the spacetime of interest, it may be that a particular set of modes 
and corresponding vacuum and Fock space emerge as having natural, 
physical meaning.  The theory of conformally-coupled fields in a 
conformally-flat spacetime is endowed with such a feature and the 
associated vacuum state is called ``conformal vacuum".  
Thus the conformal vacuum (just like the unique vacuum in field theory 
in flat spacetime) remains to be a vacuum with respect to any other 
reference frame.  This characteristic is indeed the essential advantage 
that greatly simplifies the task of quantizing fields conformally-coupled 
to a conformally-flat spacetime. Now, in order to discuss the 
many-particle interpretation of the canonical quantization programme 
via Hamiltonian and momentum operators, we begin with the energy-momentum 
tensor (or stress tensor) for the conformally-coupled scalar field 

\begin{eqnarray}
 T^{\mu\nu} = {2\over{\sqrt{g}}}{\delta{S}\over{\delta{g}_{\mu\nu}}} 
\end{eqnarray}
which, by a straightforward computation, turns out to be [2]

\begin{eqnarray}
 T_{\mu\nu} 
   &=& (1-2\xi)\nabla_{\mu}\Phi\nabla_{\nu}\Phi 
    + (2\xi - {1\over2})g_{\mu\nu}\nabla_{\alpha}\Phi\nabla^{\alpha}\Phi    
                                                     \nonumber   \\
  &&- 2\xi\Phi\nabla_{\mu}\nabla_{\nu}\Phi                    
    + 2\xi{g}_{\mu\nu}\Phi\Box\Phi 
    + \xi(R_{\mu\nu} - {1\over2}g_{\mu\nu}R)\Phi^2. 
\end{eqnarray}
As is well-known, generally when the classical matter action is 
invariant under the conformal transformation, the associated 
classical stress tensor is traceless.  This is indeed true 
for the case at hand since  
we are dealing with conformally-invariant scalar field theory, i.e.,

\begin{eqnarray}
 T^{\lambda}_{\lambda} 
   &=& g^{\mu\nu}T_{\mu\nu}                           \nonumber   \\         
   &=& [2(n-1)\xi + (1-{n\over2})]\nabla_{\mu}\Phi\nabla^{\mu}\Phi
    +  ({2-n\over2})[\Phi(-\Box + \xi{R})\Phi]  = 0
\end{eqnarray}
where we used $\xi = {(n-2)\over4(n-1)}$ and the Euler-Lagrange's 
equation of motion $(-\Box + \xi{R})\Phi = 0$. \\ 
Also, note that generally under the conformal transformation in eq.(5), 
the stress tensor transforms as 

\begin{eqnarray}
 T_{\mu\nu} = \Omega^{-(n-2)}\tilde{T}_{\mu\nu}          
\end{eqnarray}
where $\tilde{T}_{\mu\nu} = \tilde{T}_{\mu\nu}
(\tilde{\Phi}, \tilde{g}_{\mu\nu})$ is the stress tensor of Weyl-rescaled 
fields, $\tilde{\Phi}$ and $\tilde{g}_{\mu\nu}$.  Therefore, for the case 
of conformally-coupled scalar field in conformally-flat spacetime,  

\begin{eqnarray}
 \tilde{T}_{\mu\nu} = (\partial_{\mu}\tilde{\Phi}\partial_{\nu}\tilde{\Phi}
                    - {1\over 2}\eta_{\mu\nu}
                      \partial_{\alpha}\tilde{\Phi} 
                      \partial^{\alpha}\tilde{\Phi})
 -2\xi [\partial_{\mu}(\tilde{\Phi}\partial_{\nu}\tilde{\Phi}) - 
 \eta_{\mu\nu} \partial_{\alpha}(\tilde{\Phi}\partial^{\alpha}\tilde{\Phi})]
\end{eqnarray}
which is the stress tensor for a free, massless scalar field in flat 
spacetime plus total derivative terms.
As a consequence, the Hamiltonian and the total momentum operator becomes 
generally

\begin{eqnarray}
 H = P^{0} &=& \int d^{n-1}x\sqrt{g}\;T^{00} 
   = \int d^{n-1}x\sqrt{\tilde{g}}\;\Omega^{-2}\tilde{T}^{00}, \\
     P^{i} &=& \int d^{n-1}x\sqrt{g}\;T^{i0}
            =  \int d^{n-1}x\sqrt{\tilde{g}}\;\Omega^{-2}\tilde{T}^{i0} 
\end{eqnarray}
where we used
$P^{\mu} = \int d^{n-1}x\sqrt{g}\;T^{\mu 0}$,
$T^{\mu\nu} = g^{\mu\alpha}g^{\nu\beta}T_{\alpha\beta} 
             = \Omega^{-n-2}\tilde{T}^{\mu\nu}$.
Again, for the case of conformally-coupled scalar field in 
conformally-flat spacetime, using eqs.(14)-(16),

\begin{eqnarray}
 H   &=& \int d^{n-1}x\Omega^{-2}[{1\over2}
\{\tilde{\Pi}^2 + (\partial_{i}\tilde{\Phi})^2\} - 2\xi\partial_{i}
(\tilde{\Phi}\partial_{i}\tilde{\Phi})],  \\
 P^i &=& \int d^{n-1}x\Omega^{-2}[-\{\tilde{\Pi}(\partial_{i}\tilde{\Phi})
 - 2\xi\partial_{i}(\tilde{\Phi}\tilde{\Pi})\}].
\nonumber           
\end{eqnarray}
Now, the physical interpretation of the Hamiltonian and the momentum 
operators should be clear.  The Hamiltonian density and the momentum density 
of a conformally coupled scalar field in a conformally flat spacetime turn 
out to emerge just as a conformal factor $\Omega^{-2}(x)$ times those of 
a free scalar field in flat spacetime.  Particularly note that terms 
proportional to $\xi$ in the integrands in eq.(17), which originate from
the ``total divergence'' terms in $\tilde{T}_{\mu\nu}$ in eq.(14), cannot be
ignored in this case of conformal scalar field in conformally-coupled
spacetime due to the presence of the conformal factor $\Omega^{-2}(x)$.
Namely, due to this conformal 
factor the representation of the Hamiltonian and the momentum 
operators fails to take a simple form in terms of, say, the number  
operator.\\
Next, we provide the Feynman Green's function for this conformally-coupled 
scalar field  in conformally flat spacetime. Generally in curved spacetimes,
Green's functions for scalar fields satisfy the wave equation [2]

\begin{eqnarray}
 [-\Box_{x} + {\xi}R(x)]G_{F}(x,x') 
  = -{1\over{\sqrt{g}(x)}}\delta^{n}(x-x').             
\end{eqnarray}
And particularly, the Feynman propagator for conformally-coupled
scalar field in a conformally-flat spacetime is given by
\begin{eqnarray}
 G_{F}(x,x') 
 = [\Omega^{({{2-n}\over2})}(x)G_{F}^0(x,x')\Omega^{({{2-n}\over2})}(x')]
\end{eqnarray}
where 

\begin{eqnarray}
 G^0_{F}(x,x') = \int{{d^{n}k}\over{(2\pi)^{n}}}
                 e^{ik\cdot(x-x')}{(-1)\over k^{2}}.   \nonumber
\end{eqnarray}
is the flat, Minkowski spacetime version of the Feynman propagator 
for massless scalar fields.


\vspace*{0.5cm}

{\rm\bf 2. Quantization of conformally-coupled spinor field in 
conformally-flat spacetimes}
\\
The action for a  spinor field  conformally coupled to gravity (again 
assuming the gravity as a ``background") is given by [2]

\begin{eqnarray}
 S &=& \int d^{n}x\sqrt{g}\{{i\over2}
       [\bar{\Psi}\gamma^{\mu}(\nabla_{\mu}\Psi) 
     - (\nabla_{\mu}\bar{\Psi}){\gamma^{\mu}\Psi}]\}     \nonumber  \\
   &=& \int d^{n}x\sqrt{g}\{{i\over2}
       [\bar{\Psi}\gamma^{a}e^{\mu}_{a}(x)(\nabla_{\mu}\Psi) 
     - (\nabla_{\mu}\bar{\Psi}){\gamma^{a}e^{\mu}_{a}(x)\Psi}]\}     
\end{eqnarray}
where $e^{\mu}_{a}(x)$ is the $\it n$-bein that can be considered as the 
square root of the metric 
$g_{\mu\nu}(x) = \eta_{ab}e_{\mu}^{a}(x)e_{\nu}^{b}(x)$, with
$e_{\mu}^{a}e_{b}^{\mu} = \delta^{a}_{b}$, 
$e_{a}^{\mu}e_{\nu}^{a} = \delta^{\mu}_{\nu}$ and thus 
$\gamma^{\mu}(x) = e^{\mu}_{a}(x)\gamma^{a} $ is the curved spacetime 
$\gamma$-matrices obeying 
$\{\gamma^{\mu}(x) , \gamma^{\nu}(x)\} = -2g^{\mu\nu}(x) $.
$\nabla_{\mu}(x)\equiv[\partial_{\mu} 
- {i\over4}\omega^{ab}_{\mu}(x)\sigma_{ab}]$ is the covariant derivative 
with $\omega_{\mu}^{ab}(x)$ being the spin connection and 
$\sigma_{ab} = {i\over2}[\gamma_{a} , \gamma_{b}]$ being the $SO(n-1,1)$ 
group generator in the spinor representation.\\  
Again, one can readily check that this action is indeed invariant under 
the Weyl-rescaling

\begin{eqnarray}
 g_{\mu\nu} &=& \Omega^{2}(x)\tilde{g}_{\mu\nu}, \qquad {\rm or} \qquad \ \
 e^{a}_{\mu} =  \Omega(x)\tilde{e}^{a}_{\mu},     \qquad \qquad
 e_{b}^{\mu} =  \Omega^{-1}(x)\tilde{e}_{b}^{\mu},    \nonumber \\
 \Psi(x)    &=& \Omega^{-({{n-1}\over2})}\tilde{\Psi}(x),  \\
 \bar{\Psi }(x) &=& \Omega^{-({{n-1}\over2})}\bar{\tilde{\Psi}}(x)
\nonumber
\end{eqnarray}
In particular, if the background spacetime is ``conformally-flat", 
then from $\tilde{g}_{\mu\nu} = \eta_{\mu\nu}$, it follows that 
$\tilde{e}_{\mu}^{a} = \delta_{\mu}^{a}$, 
$\tilde{e}^{\mu}_{b} = \delta^{\mu}_{b}$, 
$\tilde{\omega}^{a}_{\mu b} = 0$ and hence 
$\tilde{\nabla}_{\mu} = \partial_{\mu}$ leaving us with the action 

\begin{eqnarray}
  S = \int d^{n}x\{{i\over 2}
      {[\bar{\tilde{\Psi}}\gamma^{\mu}(\partial_{\mu}\tilde{\Psi})
    - (\partial_{\mu}\bar{\tilde{\Psi}}){\gamma^{\mu}\tilde{\Psi}}]}\}      
\end{eqnarray}
Like in the case of conformally-coupled scalar field, this implies that 
the theory of conformally-coupled spinor field in conformally-flat 
spacetime can be  substituted with the usual free, massless 
spinor field theory in flat spacetime of the Weyl-rescaled field,
$\tilde{\Psi}(x) = \Omega^{({{n-1}\over2})}(x)\Psi(x)$. Thus in order to 
eventually quantize the original field $\Psi(x)$, we first carry out 
the familiar quantisation programme for the free, Weyl-rescaled field 
$\tilde{\Psi}(x)$ and then use its result to recover the quantized theory 
of the original field 
$\Psi(x)$ via $\Psi(x) = \Omega^{-({{n-1}\over2})}(x)\tilde{\Psi}(x)$ 
at the end.  
Again, if we adopt the canonical quantisation scheme, 
we begin, for the particle interpretation, by ``mode expanding" the 
spinor field in terms of the mode functions which are the spinors 
satisfying Dirac equations
$i\gamma^{\mu}\vec{\partial}_{\mu}\tilde{\Psi} = 0,
 \bar{\tilde{\Psi}}i\gamma^{\nu}\overleftarrow{\partial}_{\nu} = 0,$
\begin{eqnarray}
 \tilde{\Psi}(x) 
    &=& \int{{d^{n-1}p\over(2\pi)^{n-1}2p^{0}}}\sum\limits_{s}
        [b(p,s)\tilde{u}(p,s)e^{ip{\cdot}x} 
       + d^{\dagger}(p,s)\tilde{v}(p,s)e^{-ip{\cdot}x}],       \\  
 \bar{\tilde{\Psi}}(x) 
    &=& \int{{d^{n-1}p\over{(2\pi)^{n-1}2p^{0}}}}\sum\limits_{s}
        [b^{\dagger}(p,s)\bar{\tilde{u}}(p,s)e^{-ip{\cdot}x}
       + d(p,s)\bar{\tilde{v}}(p,s)e^{ip{\cdot}x}]       
\end{eqnarray}
where $b (d)$ is the positive energy particle (antiparticle) annihilation 
operator and $b^{\dagger}$ $(d^{\dagger})$ is its creation operator.\\  
Now, slightly differently from the quantisation of bosons (such as the 
scalar field discussed earlier), we first demand ``anticommutation 
relations" to creation and annihilation operators to arrive eventually at 
the Pauli exclusion principle
and from them, next we determine the anticommutation relations between 
fields via the mode expansion given above in eqs.(23) and (24)  

\begin{eqnarray}
\{\tilde{\Psi}_{\alpha}(t,\vec{x}), \tilde{\Psi}^{\dagger}_{\beta}(t,\vec{y})\}
 &=& \delta_{\alpha\beta}\delta^{n-1}(\vec{x} - \vec{y}),  \\
\{\tilde{\Psi}_{\alpha}(t,\vec{x}) , \tilde{\Psi}_{\beta}(t,\vec{y})\} 
 &=& \{\tilde{\Psi}^{\dagger}_{\alpha}(t,\vec{x}) , 
       \tilde{\Psi}^{\dagger}_{\beta}(t,\vec {y})\}
  = 0 \nonumber
\end{eqnarray}
where the momentum conjugate to the Weyl-rescaled spinor field is given by
$\tilde{\Pi}(x) = i\tilde{\Psi}^{\dagger}(x)$.
Here, the essential reason for demanding anticommutation relations among 
the creation and annihilation operators is to obtain the correct 
``positive definite" total Hamiltonian operator as will be shown below.  
Now, the mode expansion of the original field is recovered as 

\begin{eqnarray}
 \Psi (x) 
  = \Omega^{-({{n-1}\over2})}(x)                 
  \int {{d^{n-1}p\over(2\pi)^{n-1}2p^{0}}}\sum\limits_{s}
      [b(p,s)\tilde{u}(p,s)e^{ip{\cdot}x}
     + d^{\dagger}(p,s)\tilde{v}(p,s)e^{-ip{\cdot}x}]  
\end{eqnarray}
and similarly for the adjoint spinor $\bar{\Psi}(x)$.      
Of course, the vacuum state associated with these mode functions of the 
original field above, namely, $b(p,s)|0> = 0$, $d(p,s)|0> = 0$ is the 
``conformal vacuum".\\ 
Next, in order to construct eventually the Hamiltonian and momentum 
operators, we again consider the stress tensor for the 
conformally-coupled spinor field [2]
\begin{eqnarray}
 T^{\mu\nu} &=& {1\over 2(det ~e)}\eta^{ab}
                [e^{\mu}_{a}{{\delta{S}}\over{\delta{e}_{\nu}^{b}}}
               + e^{\nu}_{b}{{\delta{S}}\over{\delta{e}_{\mu}^{a}}}] 
   \nonumber  \\
            &=& {i\over4}\{
                [\bar{\Psi}\gamma^{\mu}(\nabla^{\nu}\Psi)
              - (\nabla^{\nu}\bar{\Psi})\gamma^{\mu}\Psi]
              + [\bar{\Psi}\gamma^{\nu}(\nabla^{\mu}\Psi)
              - (\nabla^{\mu}\bar{\Psi})\gamma^{\nu}\Psi]\}.
\end{eqnarray}
Once again, since we are dealing with conformally-invariant spinor field 
theory,  the classical stress tensor above should be traceless as can be 
easily seen as follows 

\begin{eqnarray}
 T^{\lambda}_{\lambda} = g_{\mu\nu}T^{\mu\nu} = {i\over 2}
                 [\bar{\Psi}\gamma^{\mu}(\nabla_{\mu}\Psi)
               - (\nabla_{\mu}\bar{\Psi})\gamma^{\mu}\Psi] = 0
\end{eqnarray}
where we used the on-shell condition, 
$i\gamma^{\mu}\overrightarrow{\nabla}_{\mu}\Psi = 0$, 
$\bar{\Psi}i\gamma^{\mu}\overleftarrow{\nabla}_{\mu} = 0$.\\          
Note also that under the conformal transformation in eq.(21), 
the stress tensor above transforms as 

\begin{eqnarray}
 T_{\mu\nu} = \Omega^{-(n-2)}\tilde{T}_{\mu\nu}            
\end{eqnarray}
where $\tilde{T}_{\mu\nu} = \tilde{T}_{\mu\nu}(\tilde{\Psi} , 
\tilde{g}_{\mu\nu})$ is the stress tensor of 
Weyl-rescaled fields $\tilde{\Psi}$ and $\tilde{g}_{\mu\nu}$.
Therefore, for the case of conformally-coupled spinor field in 
conformally-flat spacetime  

\begin{eqnarray}
   \tilde {T}_{\mu\nu} = {i\over4}
    \{[\bar{\tilde{\Psi}}\gamma_{\mu}(\partial_{\nu}\tilde{\Psi})
       - (\partial_{\nu}\bar{\tilde{\Psi}})\gamma_{\mu}\tilde{\Psi}]
    + [\bar{\tilde{\Psi}}\gamma_{\nu}(\partial_{\mu}\tilde{\Psi})
       - (\partial_{\mu}\bar{\tilde{\Psi}})\gamma_{\nu}\tilde{\Psi}]\}
\end{eqnarray}
which is just the stress tensor for free, massless spinor in flat 
spacetime.\\
As a consequence, the Hamiltonian and the total momentum operator become 
generally

\begin{eqnarray}
  H = P^{0} &=& \int d^{n-1}x\sqrt{g}\;T^{00} 
    = \int d^{n-1}x\sqrt{\tilde{g}}\;\Omega^{-2}\tilde{T}^{00}, \\ 
 P^{i} &=& \int d^{n-1}x\sqrt{g}\;T^{i0} 
        = \int d^{n-1}x\sqrt{\tilde{g}}\;\Omega^{-2}\tilde{T}^{i0} 
\end{eqnarray}
where we used
$P^{\mu}    = \int d^{n-1}x\sqrt{g}\;T^{\mu 0}$ as before and                
$T^{\mu\nu} = g^{\mu\alpha}g^{\nu\beta}T_{\alpha\beta} 
             = \Omega^{-n-2}\tilde{T}^{\mu\nu}$.          
Again, for the case of conformally-coupled spinor field in 
conformally-flat spacetime, using eq.(46)

\begin{eqnarray}
 H &=& \int d^{n-1}x \Omega^{-2}{i\over 2}
       [\bar{\tilde{\Psi}}\gamma_{0}(\partial_t\tilde{\Psi})
        - (\partial_t\bar{\tilde{\Psi}})\gamma_{0}\tilde{\Psi}],\\
 P^i &=& \int d^{n-1}x \Omega^{-2}{-i\over 4} 
   \{ [\bar{\tilde{\Psi}}\gamma_{i}(\partial_{t}\tilde{\Psi})
       - (\partial_{t}\bar{\tilde{\Psi}})\gamma_{i}\tilde{\Psi}]
    + [\bar{\tilde{\Psi}}\gamma_{0}(\partial_{i}\tilde{\Psi})
       - (\partial_{i}\bar{\tilde{\Psi}})\gamma_{0}\tilde{\Psi}]\}     
\end{eqnarray}
Again, the physical interpretation of the Hamiltonian and the momentum 
operators is clear.  The Hamiltonian density and the momentum density of 
a conformally coupled spinor field in a conformally flat spacetime emerge 
as a conformal factor $\Omega^{-2}(x)$ times those of a free spinor field 
in flat spacetime.  And as stated earlier, this conformal factor keeps us 
from representing the Hamiltonian and the momentum operators in a simple 
form in terms of a number operator.\\ 
Next, we provide the Feynman Green's function for this conformally-coupled 
spinor field in conformally flat spacetime. Generally in curved spacetimes,
Green's functions for the massless spinor field satisfies the wave equation
[2]
\begin{eqnarray}
 [i\gamma^{\mu}\nabla^{x}_{\mu}]S_{F}(x,x')
  = {1\over {\sqrt{g(x)}}}\delta^n(x-x').        
\end{eqnarray}
And particularly, the Feynman propagator for conformally-coupled spinor
field in a conformally-flat spacetime is given by
\begin{eqnarray}
 S_{F}(x,x') = \Omega^{({{1-n}\over 2})}(x)
               S^{0}_{F}(x,x')
               \Omega^{({{1-n}\over 2})}(x')      
\end{eqnarray}
where

\begin{eqnarray}
 S^0_{F}(x,x') = \int {{d^{n}p}\over {(2\pi)^n}}
                 e^{ip\cdot (x-x')}
                 {{p_{\mu}\gamma^{\mu}}\over {p^2 + i\epsilon}}   
\end{eqnarray}
is the flat, Minkowski spacetime version of the Feynman propagator
for massless spinor fields.
For future use, we note that the relationship between the Dirac propagator 
above and the Klein-Gordon propagator (generally in the presence 
of the mass term)
\begin{eqnarray}
G^{0}_{F}(x,x') =\int{{d^{n}p}\over{(2\pi)^n}}e^{ip{\cdot}(x-x')} 
{{(-1)}\over{p^2 + m^2}} \nonumber 
\end{eqnarray}
in flat spacetime is given by 

\begin{eqnarray}
 S^{0}_{F}(x,x') = [i\gamma^{\alpha}\partial^{x}_{\alpha} + m]
                       G^{0}_{F}(x,x').
\end{eqnarray}
Later on, we shall use this relation to obtain the Green's functions
for conformally-coupled spinor field from those for conformally-coupled
scalar field.
\vspace*{0.5cm}

\noindent
\begin{center}
{\rm\bf III. Quantization in the background of AdS$_3$ black hole spacetime}
\end{center}
{\rm\bf 1. Introducing ``conformal gauge''}    
\\
Recently, a stationary, axisymmetric solution to the AdS$_3$
Einstein equation $R_{\mu \nu} - {1\over 2}g_{\mu \nu}R + 
\Lambda g_{\mu \nu} = 0$ that can be interpreted as a spinning
black hole solution has been discovered by Banados, Teitelboim
and Zanelli (BTZ) [3]. 
The $\rm AdS_3$ black hole solution in Schwarzschild-type coordinates 
$x^{\mu} = ( t,r,\phi ) $ can be given
in Arnowitt-Deser-Misner's (ADM) (2+1) space-plus-time split form
by [3,4]
\begin{eqnarray}
 ds^2 &=& g_{\mu\nu}dx^{\mu}dx^{\nu} 
                                                    \nonumber  \\
      &=& - (N^{\perp})^2dt^2 + f^{-2}dr^2 + r^2(d\phi + N^{\phi}dt)^2  \\
               {\rm where}    \qquad
 {N^\perp} 
      &=& f =\Bigl(- M + {r^2\over l^2} + {J^2\over{4r^2}}\Bigr)^{1\over2} 
           , \qquad N^{\phi} = - {J\over{2r^2}}    \nonumber
\end{eqnarray}
are the lapse and the angular shift respectively. Of course the
$\phi$-coordinate here is taken to be a periodic, angular coordinate
satisfying $\phi = \phi + 2\pi n$ $(n \in Z)$. The parameter
$l$ is related to the negative cosmological constant by
$l^{-2} = - \Lambda $ and $M$ and $a = J/2$ are the ADM mass and the
angular momentum per unit mass of the hole respectively.
Perhaps it would be appropriate to give a brief description of the 
structure of this black hole spacetime. The causal structure of this
spinning AdS$_3$ black hole has, in many ways, close similarity to
that of Kerr black hole spacetime in 4-dim. To begin, since this
black hole solution is stationary and axisymmetric, it possesses
two Killing fields $\xi^{\mu} = (\partial /\partial t)^{\mu}$ and
$\psi^{\mu} = (\partial /\partial \phi)^{\mu}$ correspondingly.
And it is their linear combination $\chi^{\mu} = \xi^{\mu} +
\Omega_{H}\psi^{\mu}$ [10] which is normal to the 
Killing horizons of this 
spinning hole. Normally, this is the defining equation of the 
angular velocity $\Omega_{H}$ of the rotating holes. 
Now the spinning AdS$_3$ black hole solution given in eq.(39) has
regular Killing horizons at points where the Killing field
$\chi^{\mu}$ becomes null, namely at
\begin{eqnarray}
r_{\pm}
      = l\;\Biggl[
            {M\over 2}\Bigl\{1\pm\sqrt{1-({J\over Ml})^2}\Bigr\}
           \Biggr]^{1\over 2}                      
\end{eqnarray}
(i.e., $M = (r^2_{+}+r^2_{-})/l^2$ and $a = r_{+}r_{-}/l$) with
$r_{+}$ being the event horizon and $r_{-}$ the inner Cauchy horizon
respectively provided $\mid a\mid \leq Ml/2$. The angular velocity 
of the event horizon, then, can be evaluated as
\begin{eqnarray}
\Omega_{H} = -{g_{t\phi}\over g_{\phi \phi}}\mid_{r_{+}} =
{a\over r^2_{+}}.
\end{eqnarray}
This black hole, as one may expects, possesses the ``ergoregion''
as well. And the ``static limit'', i.e., the outer boundary of
this ergoregion
occurs at $r_{s} = \sqrt{M}l > r_{+}$. As we shall discuss later on, the
existence of the ergoregion in this black hole spacetime naturally
stimulates our curiosity concerning the possible superradiant
scattering phenomenon which is known to occur in 4-dim. situation.
Having reviewed the structure of the AdS$_3$ black hole spacetime, 
now consider a coordinate transformation of the AdS$_3$ black hole 
metric to the ``conformal gauge''[4]
\begin{eqnarray}
 (t , r , \phi) \rightarrow (z_0 , z_1 , z_2). \nonumber
\end{eqnarray}
Since there are two coordinate singularities at $r=r_{+}$ and
$r_{-}$, we need two coordinate patches one defined around $r_{+}$
and the other around $r_{-}$ (as we usually do when transforming
to Kruskal-type coordinates). \\
(1) Coordinate patch A around the event horizon at $r=r_{+}$ : \\
For $r\ge r_{+}\qquad \qquad $            (Region I)

\begin{eqnarray}
  z_0 
  &=& \Bigl({{r^2 - r^2_{+}}\over{r^2 - r^2_{-}}}  \Bigr)^{1\over2}\,
  \sinh \Bigl({r_{+}\over l^2}t  - {r_{-}\over l}\phi\Bigr)\,
  \exp{\Bigl[{r_{+}\over l}\phi - {r_{-}\over l^2}t \Bigr]},      
                                                   \nonumber  \\
  z_1 
  &=& \Bigl({{r^2 - r^2_{+}}\over{r^2 - r^2_{-}}}  \Bigr)^{1\over 2}\,
  \cosh \Bigl({r_{+}\over l^2}t  - {r_{-}\over l}\phi\Bigr) \,
  \exp{\Bigl[{r_{+}\over l}\phi - {r_{-}\over l^2}t \Bigr]},      
                                                   \nonumber  \\
  z_2 
  &=& \Bigl({{r^2_{+} - r^2_{-}}\over {r^2 - r^2_{-}}}\Bigr)^{1\over 2}\,
  \exp{\Bigl[{r_{+}\over l}\phi - {r_{-}\over l^2}t\Bigr]}.      
                                                   \nonumber  
\end{eqnarray}
For $r_{-} < r \le r_{+} \qquad $   (Region II)

\begin{eqnarray}
  z_0 
  &=& \Bigl({{r^2_{+} - r^2}\over{r^2 - r^2_{-}}}  \Bigr)^{1\over 2}\,
  \cosh \Bigl({r_{+}\over l^2}t  - {r_{-}\over l}\phi\Bigr)\,
  \exp{\Bigl[{r_{+}\over l}\phi - {r_{-}\over l^2}t \Bigr]},      
                                                   \nonumber  \\
  z_1 
  &=& \Bigl({{r^2_{+} - r^2}\over{r^2 - r^2_{-}}}  \Bigr)^{1\over 2}\,
  \sinh \Bigl({r_{+}\over l^2}t  - {r_{-}\over l}\phi\Bigr)\,
  \exp{\Bigl[{r_{+}\over l}\phi - {r_{-}\over l^2}t \Bigr]},      
                                                              \\
  z_2 
  &=& \Bigl({{r^2_{+} - r^2_{-}}\over {r^2 - r^2_{-}}}\Bigr)^{1\over 2}\,
  \exp{\Bigl[{r_{+}\over l}\phi - {r_{-}\over l^2}t\Bigr]}.      
                                                   \nonumber
\end{eqnarray}
(2) Coordinate patch B around the Cauchy horizon at $r=r_{-}$ : \\
For $r_{-} \le r < r_{+} \qquad $   (Region II)

\begin{eqnarray}
  z_0
  &=& \Bigl({{r^2 - r^2_{-}}\over{r^2_{+} - r^2}}  \Bigr)^{1\over 2}\,
  \cosh \Bigl({r_{-}\over l^2}t  - {r_{+}\over l}\phi\Bigr)\,
  \exp{\Bigl[{r_{-}\over l}\phi - {r_{+}\over l^2}t \Bigr]},
                                                   \nonumber  \\
  z_1
  &=& \Bigl({{r^2 - r^2_{-}}\over{r^2_{+} - r^2}}  \Bigr)^{1\over 2}\,
  \sinh \Bigl({r_{-}\over l^2}t  - {r_{+}\over l}\phi\Bigr)\,
  \exp{\Bigl[{r_{-}\over l}\phi - {r_{+}\over l^2}t \Bigr]},
                                                              \\
  z_2
  &=& \Bigl({{r^2_{+} - r^2_{-}}\over {r^2_{+} - r^2}}\Bigr)^{1\over 2}\,
  \exp{\Bigl[{r_{-}\over l}\phi - {r_{+}\over l^2}t\Bigr]}.
                                                   \nonumber
\end{eqnarray}
For ${\rm 0} < r \le r_{-} \qquad $   (Region III)

\begin{eqnarray}
  z_0 
  &=& \Bigl({{r^2_{-} - r^2}\over{r^2_{+} - r^2}}  \Bigr)^{1\over 2}\,
  \sinh \Bigl({r_{-}\over l^2}t  - {r_{+}\over l}\phi\Bigr)\,
   \exp{\Bigl[{r_{-}\over l}\phi - {r_{+}\over l^2}t \Bigr]},      
                                                    \nonumber  \\
  z_1 
  &=& \Bigl({{r^2_{-} - r^2}\over{r^2_{+} - r^2}}  \Bigr)^{1\over 2}\,
   \cosh \Bigl({r_{-}\over l^2}t  - {r_{+}\over l}\phi\Bigr)\,
   \exp{\Bigl[{r_{-}\over l}\phi - {r_{+}\over l^2}t \Bigr]},      
                                                    \nonumber  \\
  z_2 
  &=& \Bigl({{r^2_{+} - r^2_{-}}\over {r^2_{+} - r^2}}\Bigr)^{1\over 2}\,
   \exp{\Bigl[{r_{-}\over l}\phi - {r_{+}\over l^2}t\Bigr]}      
                                                    \nonumber
\end{eqnarray}
in terms of which the $\rm AdS_3$ black hole metric takes 
the ``conformally flat" form [4]

\begin{eqnarray}
 ds^2 = {l^2\over z_2^2}(- dz^2_0 + dz^2_1 + dz^2_2)
      = \Omega^2(z)\eta_{\mu\nu}dz^{\mu}dz^{\nu}.   
\end{eqnarray}
Note that this type of coordinate transformation is essentially
of the same kind as the transformation from Schwarzschild-type to
Kruskal coordinates in, say, Reissner-Nordstrom black hole
spacetime in 4-dim. Namely, by patching two charts, one
well-defined only for all $r > r_{-}$ and the other again 
well-defined only for all $r < r_{+}$, one can go over to a
new coordinate system $z^{\mu} = (z_{0}, z_{1}, z_{2})$ which
is free of coordinate singularities and can cover the whole
of black hole spacetime, not just a part of it.
And of course in order for these coordinatizations, 
(both $(t, r, \phi)$ and $(z_{0}, z_{1}, z_{2})$) to parametrize 
a black hole spacetime, not just the universal covering space of
AdS$_3$ (i.e., CAdS$_3$), an extra condition needs to be imposed.
Namely, since the BTZ
black hole solution can be obtained from CAdS$_3$ via discrete
identifications of points, i.e., the action of discrete
subgroup of the AdS$_3$ isometry group $SO(2,2)$, in order
for these coordinatizations to represent the BTZ black hole
spacetime, we should implicitly assume the identifications
\begin{eqnarray}
\phi = \phi + 2\pi n    ~~~~(n\in Z). \nonumber
\end{eqnarray}
As is manifest in this conformal gauge, this BTZ black
hole spacetime is indeed conformally-flat as we stressed
in the introduction. Thus if one considers
quantum fields coupled conformally to this conformally-flat
background spacetime, computations of quantities like the
mode expansion forms and two-point Green's functions can be
carried out exactly.
The region II, which lies between the two horizons, is the overlap
of two coordinate patches A and B. Thus in order to describe
this overlap region, one may select any coordinates belonging to
the patch A or to the patch B. Here in this work, we shall take
the coordinates belonging to patch A in constructing mode
expansion forms (given in the following subsection) and Green's 
functions (which will be given in the appendix). 
Note that for the sake of notational simplicity, we shall henceforth
use the redefined parameters $a_{\pm} \equiv r_{\pm}/l^2$. 


\vspace*{0.5cm}

{\rm\bf 2. Mode expansion of conformally-coupled  fields in 
AdS$_3$ black hole spacetime}                                

\noindent
{\it i) Real scalar field}
\\ 
According to the general formulation discussed in the preceding section,
the mode expansion for a conformally-coupled real scalar field in 
3-dim. conformally-flat spacetime is

\begin{eqnarray}
\Phi(x) = \Omega^{-{1\over 2}}(x)\int {d^2k\over (2\pi)^{2}2\omega_{k}}
        \Bigl[a(k)e^{ik\cdot x} + a^{\dagger}(k)e^{-ik\cdot x}\Bigr]  
                                                      \nonumber
\end{eqnarray}
Thus,  for the case of our AdS$_3$ black hole background

\begin{eqnarray}
 {\Phi}(z) = {({l\over {z_2}})}^{-{1\over 2}}
             \int {{d^2\!k}\over (2\pi)^2 2\omega_{k}}
             \Bigl[
               a(k)e^{ik_{\mu}z^{\mu}} + a^{\dagger}(k)e^{-ik_{\mu}z^{\mu}}
             \Bigr].
\end{eqnarray}
where $z^{\mu}=(z_0, z_1, z_2)$ is to be substituted by the original coordinates
$(t, r, \phi)$ via the coordinate transformation laws given earlier. \\
Here, recall that in the coordinatization $(t, r, \phi)$ for the BTZ black
hole, the identification $\phi = \phi + 2\pi m$ $(m \in Z)$ was implicitly understood.
Thus both the mode expansion forms of the field operators being discussed in
this section and the Green's functions that will be provided later in the appendix
should exhibit the periodicity in $\phi$-coordinate, i.e.,
$\Phi (t, r, \phi) = \Phi (t, r, \phi+2\pi m$). This can be achieved by taking
the infinite linear sum
\begin{eqnarray}
\Phi (t, r, \phi) = \sum_{n=-\infty}^{\infty} \Phi (t, r, \phi_{n}) \nonumber
\end{eqnarray}
where $\phi_{n} = \phi + 2\pi n$. Then by the superposition principle, this 
infinite linear sum is automatically a solution to the field equation as well
since each $ \Phi (t, r, \phi_{n})$ is a solution as one can check in a
straightforward manner. As we shall see in the appendix, Green's functions
exhibiting the same periodicity in $\phi$ can also be constructed via the 
same infinite sum
\begin{eqnarray}
G^{+}(\Delta t, \Delta r, \Delta \phi) = 
\sum_{n=-\infty}^{\infty} G^{+}(\Delta t, \Delta r, \Delta \phi_{n}) \nonumber
\end{eqnarray}
with $\Delta \phi_{n} = (\phi - \phi')+2\pi n$. And this procedure for 
constructing the Green's function amounts to employing the method of images [15]. \\
For $ r\ge r_{+}\qquad \qquad $  (Region I)

\begin{eqnarray}
 &\Phi & ( t,r,\phi_{n}) = l^{-{1\over 2}}
      \Bigl({{r^2_{+} - r^2_{-}}\over {r^2 - r^2_{-}}}\Bigr)^{1\over 4}
      e^{{1\over 2}(la_{+}\phi_{n} - a_{-}t)}
      \int {{d^2\!k}\over (2\pi)^2 2\omega_{k}}
      \Bigl[a(k)e^{iE_{I}} + a^{\dagger}(k)e^{-iE_{I}}\Bigr]    
                                                        \nonumber    \\
      {\rm with }                                       \nonumber    \\
 &E&_{I} = e^{(la_{+}\phi_{n} - a_{-}t)}                    
       \Bigl[({{r^2 - r^2_{+}}\over 
              {r^2 - r^2_{-}}})^{1\over 2}
             \{ - k^{0}\sinh (a_{+}t -la_{-}\phi_{n})    
           + k^{1}\cosh (a_{+}t - la_{-}\phi_{n})\}         
             + ({{r^2_{+} - r^2_{-}}\over 
              {r^2 - r^2_{-}}})^{1\over 2}k^2
         \Bigr].                                        \nonumber  
\end{eqnarray}
For $ r_{-} < r \le r_{+}\qquad $  (Region II)
 
\begin{eqnarray}
 &\Phi & ( t,r,\phi_{n}) = l^{-{1\over 2}}
      \Bigl({{r^2_{+} - r^2_{-}}\over{r^2 - r^2_{-}}}\Bigr)^{1\over 4}
      e^{{1\over 2}(la_{+}\phi_{n} - a_{-}t)}
      \int {{d^2\!k}\over (2\pi)^2 2\omega_{k}}
      \Bigl[a(k)e^{iE_{II}} + a^{\dagger}(k)e^{-iE_{II}}\Bigr]     
                                                        \nonumber    \\
      {\rm with }                                       \nonumber    \\
 &E&_{II} = e^{(la_{+}\phi_{n} - a_{-}t)}                   
         \Bigl[({{r^2_{+} - r^2}\over 
            {r^2 - r^2_{-}}})^{1\over 2}
            \{ - k^{0}\cosh (a_{+}t -la_{-}\phi_{n})     
            + k^{1}\sinh (a_{+}t - la_{-}\phi_{n}) \}
               + ({{r^2_{+} - r^2_{-}}\over 
             {r^2 - r^2_{-}}})^{1\over 2}k^2
            \Bigr].
                                                        \nonumber
\end{eqnarray}
For $ {\rm 0} < r \le r_{-} \qquad $ (Region III)
      
\begin{eqnarray}
 &\Phi & ( t,r,\phi_{n}) = l^{-{1\over 2}}
      \Bigl({{r^2_{+} - r^2_{-}}\over{r^2_{+} - r^2}}\Bigr)^{1\over 4}
      e^{{1\over 2}(la_{-}\phi_{n} - a_{+}t)}
      \int {{d^2\!k}\over (2\pi)^2 2\omega_{k}}
      \Bigl[a(k)e^{iE_{III}} + a^{\dagger}(k)e^{-iE_{III}}\Bigr]     
                                                       \nonumber    \\
      {\rm with }                                      \nonumber    \\
 &E&_{III} = e^{(la_{-}\phi_{n} - a_{+}t)}              
          \Bigl[({{r^2_{-} - r^2}\over 
             {r^2_{+} - r^2}})^{1\over 2}
             \{ - k^{0}\sinh (a_{-}t - la_{+}\phi_{n})  
             + k^{1}\cosh (a_{-}t - la_{+}\phi_{n}) \}
                + ({{r^2_{+} - r^2_{-}}\over 
             {r^2_{+} - r^2}})^{1\over 2}k^2
             \Bigr]
                                                       \nonumber
\end{eqnarray}
where $k^{0}=[(k^{1})^2 + (k^{2})^2]^{1/2}$.
Despite the general formulation we provided in the sect.II, suspicious
readers may wonder if the mode expansions obtained in this way are
really solutions to the Klein-Gordon equation. Thus we ensure that
we checked by a straightforward calculation that the mode expansions
above do satisfy the Klein-Gordon equation in AdS$_3$ black hole
spacetime written in the Schwarzschild-type coordinates.


\vspace*{0.5cm}

\noindent
{\it ii)  Complex scalar field}                    
\\            
Generally in $\it n$-dimensional conformally-flat spacetime, the
mode expansion of conformally-coupled complex scalar field is
given by
\begin{eqnarray}
 \Phi(x) =     \Omega^{-{({{n-2}\over 2})}}(x)\tilde{\Phi}(x)  , \qquad
 \Phi^{*}(x) = \Omega^{-{({{n-2}\over 2})}}(x)\tilde{\Phi^{*}}(x)
 \nonumber 
\end{eqnarray}
with
\begin{eqnarray}
 \tilde{\Phi}(x) 
    &=& \int {{d^{n-1}k}\over {(2\pi)^{n-1}2\omega_{k}}}
        [a(k)e^{ik\cdot x} + b^{\dagger}(k)e^{-ik\cdot x}], \nonumber \\
 \tilde{\Phi}^{*}(x) 
    &=& \int {{d^{n-1}k}\over {(2\pi)^{n-1}2\omega_{k}}}
        [b(k)e^{ik\cdot x} + a^{\dagger}(k)e^{-ik\cdot x}] \nonumber
\end{eqnarray}
where $a(k)$, $a^{\dagger}(k)$ are annihilation and creation operators for 
positive energy ``particle'' respectively, $b(k)$, $b^{\dagger}(k)$  
annihilation and creation operators for positive energy ``antiparticle''
respectively. Thus, for the case at hand, i.e., in the AdS$_3$ black hole
background

\begin{eqnarray}
 {\Phi}(z) &=& {({l\over {z_2}})}^{-{1\over 2}}
               \int {{d^2\!k}\over (2\pi)^2 2\omega_{k}}
               \Bigl[  a(k)e^{ik_{\mu}z^{\mu}} 
                     + b^{\dagger}(k)e^{-ik_{\mu}z^{\mu}}\Bigr],\\
 {\Phi^{*}}(z) &=& {({l\over {z_2}})}^{-{1\over 2}}
               \int {{d^2\!k}\over (2\pi)^2 2\omega_{k}}
               \Bigl[  b(k)e^{ik_{\mu}z^{\mu}} 
                     + a^{\dagger}(k)e^{-ik_{\mu}z^{\mu}}\Bigr].
\nonumber
\end{eqnarray}
Now the correct mode expansion is given by
\begin{eqnarray}
\Phi (t, r, \phi) = \sum_{n=-\infty}^{\infty}\Phi (t, r, \phi_{n}) \nonumber
\end{eqnarray}
and similarly for $\Phi^{*}(t, r, \phi)$. \\
For $ r\ge r_{+}\qquad $  (Region I)

\begin{eqnarray}
 &\Phi & ( t,r,\phi_{n}) = l^{-{1\over 2}}
      \Bigl({{r^2_{+} - r^2_{-}}\over {r^2 - r^2_{-}}}\Bigr)^{1\over 4}
      e^{{1\over 2}(la_{+}\phi_{n} - a_{-}t)}
      \int {{d^2\!k}\over (2\pi)^2 2\omega_{k}}
      \Bigl[a(k)e^{iE_{I}} + b^{\dagger}(k)e^{-iE_{I}}\Bigr],\nonumber \\    
 &\Phi &^{*} ( t,r,\phi_{n}) = l^{-{1\over 2}}
      \Bigl({{r^2_{+} - r^2_{-}}\over {r^2 - r^2_{-}}}\Bigr)^{1\over 4}
      e^{{1\over 2}(la_{+}\phi_{n} - a_{-}t)}
      \int {{d^2\!k}\over (2\pi)^2 2\omega_{k}}
      \Bigl[b(k)e^{iE_{I}} + a^{\dagger}(k)e^{-iE_{I}}\Bigr]    
                                                        \nonumber    \\    
      {\rm with }                                       \nonumber    \\
 &E&_{I} = e^{(la_{+}\phi_{n} - a_{-}t)}
     \Bigl[({{r^2 - r^2_{+}}\over {r^2 - r^2_{-}}})^{1\over 2}
        \{ - k^{0}\sinh (a_{+}t -la_{-}\phi_{n})       
      + k^{1}\cosh (a_{+}t - la_{-}\phi_{n})\}
        + ({{r^2_{+} - r^2_{-}}\over {r^2 - r^2_{-}}})^{1\over 2}k^{2}
      \Bigr].
                                                        \nonumber  
\end{eqnarray}
For $ r_{-} < r \le r_{+}\qquad $  (Region II)
 
\begin{eqnarray}
 &\Phi & ( t,r,\phi_{n}) = l^{-{1\over 2}}
      \Bigl({{r^2_{+} - r^2_{-}}\over {r^2 - r^2_{-}}}\Bigr)^{1\over 4}
      e^{{1\over 2}(la_{+}\phi_{n} - a_{-}t)}
      \int {{d^2\!k}\over (2\pi)^2 2\omega_{k}}
      \Bigl[a(k)e^{iE_{II}} + b^{\dagger}(k)e^{-iE_{II}}\Bigr],     
                                                        \nonumber    \\
 &\Phi &^{*} ( t,r,\phi_{n}) = l^{-{1\over 2}}
      \Bigl({{r^2_{+} - r^2_{-}}\over {r^2 - r^2_{-}}}\Bigr)^{1\over 4}
      e^{{1\over 2}(la_{+}\phi_{n} - a_{-}t)}
      \int {{d^2\!k}\over (2\pi)^2 2\omega_{k}}
      \Bigl[b(k)e^{iE_{II}} + a^{\dagger}(k)e^{-iE_{II}}\Bigr]     
                                                        \nonumber    \\     
      {\rm with }                                       \nonumber    \\
 &E&_{II} = e^{(la_{+}\phi_{n} - a_{-}t)}                  
     \Bigl[({{r^2_{+} - r^2}\over {r^2 - r^2_{-}}})^{1\over 2}
        \{ - k^{0}\cosh (a_{+}t -la_{-}\phi_{n})                
      + k^{1}\sinh (a_{+}t - la_{-}\phi_{n}) \}
        + ({{r^2_{+} - r^2_{-}}\over {r^2 - r^2_{-}}})^{1\over 2}k^{2}
      \Bigr].
                                                        \nonumber
\end{eqnarray}
For $ {\rm 0} < r \le r_{-}\qquad $  (Region III)
      
\begin{eqnarray}
 &\Phi & ( t,r,\phi_{n}) = l^{-{1\over 2}}
      \Bigl(
        {{r^2_{+} - r^2_{-}}\over {r^2_{+} - r^2}}
      \Bigr)^{1\over 4}
      e^{{1\over 2}(la_{-}\phi_{n} - a_{+}t)}
      \int {{d^2\!k}\over (2\pi)^2 2\omega_{k}}
      \Bigl[
        a(k)e^{iE_{III}} + b^{\dagger}(k)e^{-iE_{III}}
      \Bigr],                                           \nonumber    \\
  &\Phi &^{*} ( t,r,\phi_{n}) = l^{-{1\over 2}}
      \Bigl(
        {{r^2_{+} - r^2_{-}}\over {r^2_{+} - r^2}}
      \Bigr)^{1\over 4}
      e^{{1\over 2}(la_{-}\phi_{n} - a_{+}t)}
      \int {{d^2\!k}\over (2\pi)^2 2\omega_{k}}
      \Bigl[
        b(k)e^{iE_{III}} + a^{\dagger}(k)e^{-iE_{III}}
      \Bigr]                                           \nonumber    \\     
      {\rm with }                                      \nonumber    \\
 &E&_{III} = e^{(la_{-}\phi_{n} - a_{+}t)}      
      \Bigl[({{r^2_{-} - r^2}\over {r^2_{+} - r^2}})^{1\over 2}
        \{ - k^{0}\sinh (a_{-}t - la_{+}\phi_{n})           
      + k^{1}\cosh (a_{-}t - la_{+}\phi_{n}) \}
        + ({{r^2_{+} - r^2_{-}} \over {r^2_{+} - r^2}})^{1\over 2}k^{2}
      \Bigr]                                           \nonumber
\end{eqnarray}
where again $k^{0} = [(k^{1})^2 + (k^{2})^2]^{1/2}$.
Next, consider the particle number current given by
$j^{\mu} = -i(\Phi^{\ast}\nabla^{\mu}\Phi - \Phi\nabla^{\mu}\Phi^{\ast})$
and examine the continuity, say, of the particle number density
(i.e., the pobability density) flux defined by
$n_{\mu}j^{\mu} = -\chi^{\mu}j_{\mu} = i[\Phi^{\ast}(\partial_{t}
+ \Omega_{H} \partial_{\phi})\Phi -
\Phi(\partial_{t} + \Omega_{H} \partial_{\phi})\Phi^{\ast}]$
where, as introduced earlier, 
$\chi^{\mu} = \xi^{\mu} + \Omega_{H} \psi^{\mu} = \delta^{\mu}_{t} + 
\Omega_{H} \delta^{\mu}_{\phi}$ ($\Omega_{H}$ denotes the angular velocity
of the rotating black hole) being the Killing field which is outward
normal to the outer event horizon at $r=r_{+}$ (on which $\Omega_{H} =
\Omega_{H}(r_{+})$) and to the inner null surface at  $r=r_{-}$ 
(on which $\Omega_{H} = \Omega_{H}(r_{-})$) and thus $n^{\mu}$ denotes the
inward unit vector which is opposite to the direction of $\chi^{\mu}$.
Then one can readily check that across the event horizon and the Cauchy horizon, 
$n_{\mu}j^{\mu}\mid_{r_{\pm}+\epsilon} = n_{\mu}j^{\mu}\mid_{r_{\pm}-\epsilon}$,
namely the particle number flux is conserved. The particle number
conservation across the event horizon at $r=r_{+}$ and the Cauchy horizon
at $r=r_{-}$ confirms that they are just  coordinate singularities. 


\vspace*{0.5cm}

\noindent
{\it iii) Spinor field}
\\ 
Before we construct the mode expansion of the spinor field in AdS$_3$
black hole spacetime, we should note that there are some subtle
properties of fermions in (2+1)-dim. To formulate the algebra of Dirac
matrices in (2+1)-dim., we basically need 3-anticommuting Hermitian
matrices each of whose square is unity. Clearly the standard $2\times 2$ Pauli
matrices fit the bill. Thus the minimal spinor fields we can work with
are 2-component spinors. However, since the two components should
describe positive-energy and negative-energy solutions, we are left with
just one component to describe the spin state. Actually this is indeed
enough in 2-space dimensions, where the rotation group $O(2) \sim U(1)$
is abelian and has 1-dim. irreducible representations. Or in plain 
English, the spin of a minimal fermion will point in some definite 
direction - up or down - in the particle's rest frame. Thus in the
mode expansion of minimal spinor field in (2+1)-dim. like the AdS$_3$
black hole background, the spin sum is absent.
Thus the mode expansion for a conformally-coupled spinor field in the AdS$_3$
black hole background is given by

\begin{eqnarray}
 {\Psi}(z) &=& {({l\over z_2})}^{-1}
      \int {d^2p\over (2\pi)^2 2p^0}
      \Bigl[ 
       b(p)u(p)e^{ip_{\mu}z^{\mu}} + d^{\dagger}(p)v(p)e^{-ip_{\mu}z^{\mu}} 
      \Bigr],\\
 {\bar\Psi}(z) &=& {({l\over z_2})}^{-1}
      \int {{d^2 p}\over {(2\pi)^2 2p^0}}
      \Bigl[b^{\dagger }(p){\bar u}(p)e^{-ip_{\mu}z^{\mu}}
            + d(p){\bar v}(p)e^{ip_{\mu}z^{\mu}}\Bigr]. \nonumber            
\end{eqnarray}
Then again the correct mode expansion is given by
\begin{eqnarray}
\Psi (t, r, \phi) = \sum_{n=-\infty}^{\infty} \Psi (t, r, \phi_{n}) \nonumber
\end{eqnarray}
and similarly for the adjoint spinor ${\bar\Psi}(t, r, \phi)$. \\
For $r\ge r_{+} \qquad $(Region I) 

\begin{eqnarray}
 &\Psi & ( t,r,\phi_{n} ) = l^{-1}
      \Bigl({{r^2_{+} - r^2_{-}}\over {r^2 - r^2_{-}}}
      \Bigr)^{1\over 2}
      e^{({la_{+}\phi_{n} - a_{-}t})} \int {d^2p\over (2\pi)^2 2p^0}
      \Bigl[b(p)u(p)e^{iE_{I}} + d^{\dagger}(p)v(p)e^{-iE_{I}}\Bigr]     
                                                      \nonumber    \\
      {\rm with}                                      \nonumber    \\
 &E&_{I} = e^{(la_{+}\phi_{n} - a_{-}t)}  
      \Bigl[\Bigl({{r^2 - r^2_{+}}\over {r^2 - r^2_{-}}}\Bigr)^{1\over 2}
        \{ - p^{0}\sinh (a_{+}t - la_{-}\phi_{n} )   
      + p^{1}\cosh (a_{+}t - la_{-}\phi_{n} ) \}
      + \Bigl({{r^2_{+} - r^2_{-}}\over {r^2 - r^2_{-}}}
             \Bigr)^{1\over 2}p^{2}       
        \Bigr].
                                                      \nonumber      
\end{eqnarray}
For $ r_{-} < r \le r_{+}$ (Region II)

\begin{eqnarray}
 &\Psi & ( t,r,\phi_{n} ) = l^{-1}
      \Bigl({{r^2_{+} - r^2_{-}}\over {r^2 - r^2_{-}}}
      \Bigr)^{1\over 2}
      e^{({la_{+}\phi_{n} - a_{-}t})} \int {d^2p\over (2\pi)^2 2p^0}
      \Bigl[b(p)u(p)e^{iE_{II}} + d^{\dagger}(p)v(p)e^{-iE_{II}}\Bigr]     
                                                      \nonumber      \\
      {\rm with}                                      \nonumber      \\
 &E&_{II} = e^{(la_{+}\phi_{n} - a_{-}t)}    
     \Bigl[\Bigl({{r^2_{+} - r^2}\over {r^2 - r^2_{-}}}\Bigr)^{1\over 2}
        \{ - p^{0}\cosh (a_{+}t - la_{-}\phi_{n} )  
      + p^{1}\sinh (a_{+}t - la_{-}\phi_{n} ) \}
      + \Bigl({{r^2_{+} - r^2_{-}}\over {r^2 - r^2_{-}}}
             \Bigr)^{1\over 2}p^{2} 
        \Bigr].
                                                      \nonumber
\end{eqnarray}
For ${\rm 0} < r \le r_{-}$ (Region III)

\begin{eqnarray}
 &\Psi & ( t,r,\phi_{n} ) = l^{-1}
      \Bigl({{r^2_{+} - r^2_{-}}\over {r^2_{+} - r^2}}
      \Bigr)^{1\over 2}
      e^{({la_{-}\phi_{n} - a_{+}t})} \int {d^2p\over (2\pi)^2 2p^0}
      \Bigl[b(p)u(p)e^{iE_{III}} + d^{\dagger}(p)v(p)e^{-iE_{III}}\Bigr]   
                                                      \nonumber      \\
      {\rm with}                                      \nonumber      \\
 &E&_{III} = e^{(la_{-}\phi_{n} - a_{+}t)}
     \Bigl[\Bigl({{r^2_{-} - r^2}\over {r^2_{+} - r^2}}
        \Bigr)^{1\over 2}
        \{ - p^{0}\sinh (a_{-}t - la_{+}\phi_{n} )      
      + p^{1}\cosh (a_{-}t - la_{+}\phi_{n} ) \}
      + \Bigl({{r^2_{+} - r^2_{-}}\over {r^2_{+} - r^2}}
             \Bigr)^{1\over 2}p^{2} 
        \Bigr]
                                                      \nonumber
\end{eqnarray}
and similarly for $\bar{\Psi}(t,r,\phi_{n})$. \\
We now end this subsection with some comments accounting for the relation 
between the results given above and those appeared in the recent works.
Recently, there have been attempts to determine explicit form of ``grey
body factor (or absorption coefficient)'' $\sigma_{abs}$ in the modified
Hawking radiation spectrum (or decay rate)
\begin{eqnarray}
\Gamma = {\sigma_{abs} \over e^{{\omega \over T_{H}}} - 1} \nonumber
\end{eqnarray}
of BTZ black hole. (Here, the grey body factor is defined as the ratio
of the particle flux entering the horizon to the incoming flux at
spatial infinity.) And they include the works by Birmingham, Sachs, and
Sen [17], by Lee, Kim, and Myung [18], by Teo [17], and by Dasgupta [19].
And to this end, these authors computed the mode functions of quantum
scalar or spinor fields in the BTZ black hole spacetime. The mode expansion
forms obtained there, however, appear to be different from those given here
in the present work. As a matter of fact, this is not surprising for the
following reason. In the present work, we considered only the conformally-
coupled scalar and spinor fields whereas in [17], they considered
minimally-coupled massless scalar fields and in [18], the authors considered
non-minimally coupled dilaton field in the context of low energy string theory 
in 3-dimensions. Besides in [19], the author considered conformally-coupled 
spinor field but in a rather non-standard manner. Moreover, in all of these
works, they {\it assumed} the mode functions to take the $(t, \phi)$-dependence
of the form $e^{-i\omega t + im\phi }$ (with $\omega$ and $m$ being the frequency
and the azimuthal number respectively) which, as we shall stress later on,
can be regarded as an {\it ad hoc}. Namely, this type of $(t, \phi)$-dependence
of the mode functions is by no means rigorous in that it simply has been chosen so
based on the fact that the BTZ black hole spacetime has time-translational and
rotational isometries generated by the Killing fields $(\partial /\partial t)^{\mu}$
and $(\partial /\partial \phi)^{\mu}$ respectively. In the present work, as we 
considered only the conformally-coupled fields (in the conformally-flat BTZ
black hole spacetime), we could {\it construct} the mode functions in a 
straightforward manner without having to assume any {\it ad hoc} 
$(t, \phi)$-dependence and this is why the mode functions in the present work
and those in [17-19] came out to be different. 

\vspace*{0.5cm}

\noindent

\begin{center}
{\rm\bf IV. Absence of particle creation in conformal triviality}
\end{center}

In the preceding sections, we discussed the quantisation of conformally-
coupled scalar and spinor fields in AdS$_3$ black hole spacetime which
is conformally-flat. One crucial point in this case of ``conformal
triviality'' is the absence of real particle creation.
Here in this section, we shall provide a formal argument establishing
this statement. In ``conformally-trivial'' situations [2] when conformally-
invariant fields are propagating in a conformally flat background
spacetime, due to the high degree of symmetry possessed by the spacetime,
a unique, global physical vacuum state, called conformal vacuum exists
as mentioned earlier and as a consequence no real particles are created
from the conformal vacuum. Note that the essence in Hawking's derivation 
of particle creation by black holes [13] is the non-trivial ``Bogoliubov
transformation'' [2] relation between mode functions (or the associated
creation and annihilation operators) of the field operator at early and
at late times, namely the ``mode-mixing''. Perhaps, therefore, the
most straightforward way of displaying the absence of particle production
by the present AdS$_3$ black hole spacetime is to exhibit that the
mixing of modes is absent or equivalently that the Bogoliubov coefficient
``$\beta$'' associated with the mode-mixing is zero.
Thus to do so, we write the AdS$_3$ black hole metric again in ``conformal
gauge''
\begin{eqnarray}
ds^2 = ({l\over z_2})^2 (-dz^2_{0} + dz^2_{1} + dz^2_{2})
= \Omega^2(z) \eta_{\mu\nu}dz^{\mu}dz^{\nu} \nonumber
\end{eqnarray}
then in terms of which the mode expansion form of a real scalar field
coupled conformally to this spacetime is given by
\begin{eqnarray}
\Phi(z) = ({l\over z_2})^{-1/2}\int {d^2k\over (2\pi)^2 2\omega_{k}}
[a(k)e^{ik^{\mu}z_{\mu}} + a^{\dagger}(k)e^{-ik^{\mu}z_{\mu}}] \nonumber
\end{eqnarray}
with the conformal vacuum being defined by $a(k)\mid 0> = 0$.
Consider now the mode functions
\begin{eqnarray}
u_{k}(z) = ({l\over z_{2}})^{-1/2}{1\over [(2\pi)^2 2\omega_{k}]^{1/2}}
e^{ik^{\mu}z_{\mu}} \nonumber
\end{eqnarray}
which are ``positive frequency" modes with respect to the timelike
Killing field $\partial /\partial z_{0} \equiv \partial_{0}$, i.e.,
\begin{eqnarray}
\pounds_{\partial_{0}} u_{k}(z) = -i\omega_{k} u_{k}(z).
~~~(\omega_{k} = \mid \vec{k} \mid > 0) \nonumber
\end{eqnarray}
Obviously, these modes $u_{k}(z)$, which are positive frequency with
respect to the conformal vacuum $\mid 0>$ at one time, remain so
for all time, $z_{0}$. Namely the positive frequency mode function
$u_{k}(z)$ will remain identical both at early ($z_{0} \rightarrow
-\infty$) and at late ($z_{0} \rightarrow \infty$) times and hence 
the associated conformal vacuum which was annihilated by $a(k)$ at
early times will be so at late times, too, $\mid 0,{\rm in}> =
\mid 0, {\rm out}>$. Also note that since $z_{0}$ and $t$ are
related in the region $r > r_{+}$ by the coordinate transformation 
\begin{eqnarray}
z_0 &=& ({r^2 - r^2_{+}\over r^2 - r^2_{-}})^{1/2} \sinh 
({r_{+}\over l^2}t - {r_{-}\over l}\phi)\exp [{r_{+}\over l}\phi
- {r_{-}\over l^2}t] \\
&=& ({r^2 - r^2_{+}\over r^2 - r^2_{-}})^{1/2} {1\over 2}
[\exp \{(r_{+}-r_{-})({1\over l^2}t + {1\over l}\phi)\} -
\exp \{-(r_{+}+r_{-})({1\over l^2}t - {1\over l}\phi)\}] \nonumber
\end{eqnarray}
it is clear that $t \rightarrow -\infty$ corresponds to 
$z_0 \rightarrow -\infty$ and  $t \rightarrow \infty$ corresponds
to $z_0 \rightarrow \infty$.
Therefore, in the Bogoliubov transformation
\begin{eqnarray}
a_{out}(k) &=& \sum_{k'} [\alpha_{k'k} a_{in}(k') + 
\beta^{\ast}_{k'k} a^{\dagger}_{in}(k')], \\
a^{\dagger}_{out}(k) &=& \sum_{k'} [\alpha^{\ast}_{kk'} 
a^{\dagger}_{in}(k') + \beta_{kk'} a_{in}(k')] \nonumber 
\end{eqnarray} 
the Bogoliubov coefficient $\beta_{kk'}$ associated with the mode mixing
is zero and hence there will be no particle creation, i.e.,
\begin{eqnarray}
N_{k} &=& <{\rm in}, 0\mid a^{\dagger}_{out}(k)a_{out}(k)
\mid 0, {\rm in}> \nonumber \\
&=& \sum_{k'}\mid \beta_{kk'}\mid^2 = 0. 
\end{eqnarray}
And of course the same argument holds for the exhibition of the
absence of fermionic real particle creation.
Now that we have convinced ourselves of the absence of particle creation
(i.e., both the superradiance, which is the stimulated emission and the
Hawking radiation, which is the spontaneous emission) by the AdS$_{3}$
black hole when the fields are conformally-coupled. However, one may 
still be puzzled since it has been known in the literature that
accelerating detectors (of Unruh and DeWitt) outside the event horizon
of the AdS$_3$ black hole do detect particles. Recall that according to the 
theory of particle detection formulated by Unruh [2,14] and by 
DeWitt [2,15], the
transition probability to all possible excited states when the
(accelerating) detector registered quanta is given by [2]
\begin{eqnarray}
\Gamma = c^2 \sum_{E} \mid <E \mid m(0)\mid E_{0}>\mid^2 F(\omega)
\end{eqnarray}
with $c$ being the small coupling constant, $m(\tau)$ detector's
monopole moment operator and $F(\omega)$ the ``detector response
function'' given by
\begin{eqnarray}
F(\omega) = \int_{-\infty}^{\infty}d(\Delta \tau) e^{-i\omega
\Delta \tau} g(\Delta \tau)
\end{eqnarray}
where $\omega = (E - E_{0})$ and 
$g(\Delta \tau) = G^{+}(x(\tau), x(\tau'))$ with $G^{+}(x,x')$
being the positive-frequency Wightman function.
The detector response functions have been calculated, in the
case of scalar and spinor fields coupled conformally to the
nonrotating AdS$_{3}$ black hole, by [4,5]
\begin{eqnarray}
F_{boson}(\omega) &\sim& {1\over 2}{1\over e^{\omega /T} + 1}, \\
F_{spinor}(\omega) &\sim& {i\omega\over 2}
{1\over e^{\omega /T} - 1} \nonumber
\end{eqnarray}
with the ``local temperature'' $T$ being given by
$T = r_{+}/{2\pi l (r^2-r_{+}^2)^{1/2}}$ {\it which vanishes asymptotically
(i.e., as $r \rightarrow \infty$) incidentally confirming the absence
of real particle creation}. Apparently, there also exists an issue concerning
the ``wrong statistics'' between scalar and spinor case detector response
functions known as ``statistical inversion''. We believe that the wrong
statistics can be attributed to the odd-dimensionality ($d=3$) of the
AdS$_{3}$ black hole spacetime. We are, thus, in an uncomfortable situation where
on one hand, there is no real particle creation and on the other, there
is particle detection. In order to reconcile these two seemingly conflicting
results, we need to note the distinction between the two concepts,
``particle creation'' and ``particle detection''. As a matter of fact, 
the former, particle creation by curved spacetime can be best understood
by ``Hawking effect'' [2,13], while the latter, particle detection (which is
possible even in flat Minkowski spacetime) can be best understood by
``Unruh effect'' [2]. It is clear enough that there is no particle creation
in flat Minkowski spacetime. This, however, does not necessarily mean
that particle detectors never click in this spacetime. The detection by
particle detectors depends on the state of motion of the detector.
Indeed, an accelerated detector in 2-dim. Minkowski spacetime, namely
the Rindler observer will see a non-zero particle spectrum. This is 
the phenomenon known as Unruh effect [2]. And it can be attributed to the
fact that the accelerating Rindler observers exist only in part of
the Minkowski spacetime, i.e., the ``Rindler wedge'' separated from
other regions by a horizon. Unruh effect is the well-known example of
the fact that even without particle creation, there can be particle detection.
Meanwhile in the Hawking effect [2,13], real particles are actually created
by a black hole and thus can be detected as well. \\
To conclude, our situation associated with the conformally-coupled
fields in conformally-flat AdS$_3$ black hole spacetime can be thought
of as an analogue of the Unruh effect or of the case of conformally-
coupled fields in the spatially-flat Friedmann-Robertson-Walker
spacetimes [2].  Namely, although there is no real particle creation,
particle spectrum is still detected because of the accelerating motion
of the particle detector. And as we have seen, the absence of particle
creation most clearly manifests itself in the conformal coordinates
$z^{\mu} = (z_{0}, z_{1}, z_{2})$ and the illustration of particle
detection has been done in the Schwarzschild-type coordinates
$x^{\mu} = (t, r, \phi)$ which is an accelerating coordinate system.
In view of this, it seems meaningless to consider thermodynamics of
AdS$_3$ black hole as long as quantum fields of interest are
conformally-coupled since it never radiates real particles and hence
it never possesses even a temperature (defined in the asymptotic
region, $r \rightarrow \infty$) with which to start
investigating its thermodynamic behaviors. In the presence of
non-conformal coupling such as the minimal coupling, however, the AdS$_3$ black
hole may evaporate and therefore one may consider its thermodynamics.
In this regard, it seems appropriate to check, at this point, whether
the main conclusion of the present work given above (on the absence
of real particle creation by BTZ black hole in the case of ``conformal
triviality'') can indeed be consistent with the results of recent works
[17-19] aiming at the determination of grey body factor in the decay
rate of BTZ black hole that we referred earlier. Firstly in [17,18],
the authors are considering the case when minimally-coupled scalar
field or non-minimally coupled dilaton field are employed to study
the decay (or evaporation) of BTZ black hole. As we just pointed out,
in these cases the BTZ black hole may be shown to evaporate and thus there
is no contradiction between the conclusions of the present work and the 
results of [17,18]. Secondly in [19], the author considers the case when
conformally-coupled (but in a non-standard manner) spinor field is
employed again to study the decay of the BTZ black hole, particularly to
compute the grey body factor. Since this author also employs the
conformally-coupled spinor field, it may seem that there is a contradiction
between the conclusion of this work and that of our present work.
There is, however, no contradiction. As the author of [19] mentioned
carefully in his work, ``since the BTZ metric is asymptotically 
anti-de Sitter, the local temperature measured by any timelike observer
decreases with distance and becomes zero at spatial infinity''. Thus
he chooses ``a BTZ-observer, sitting at finite radial distance, detecting
radiation''. Namely, he studies the radiation from BTZ black hole measured
by an accelerated observer placed at a ``finite'' distance from the hole. 
As we stressed, an accelerated observer can detect particle spectrum 
(``acceleration radiation'') and thus the computation in [19] does not
contradict to the absence of real particle creation by BTZ black hole
in this conformally-trivial setting. \\
In the following section, we shall demonstrate, as a concrete
evidence of the absence of real particle creation in this case of
conformal triviality, the absence of both bosonic and fermionic
superradiances, i.e., the absence of the stimulated emission. And
to do so, the explicit mode expansion forms of scalar and spinor
fields obtained in sect. III will play a central role.

\vspace*{0.5cm}

\noindent

\begin{center}
{\rm\bf V. Demonstration of the absence superradiant scatterings}
\end{center}
{\rm\bf  1. Introduction}
\\
A black hole is, by definition, a ``region of no escape". No massive object
or even the massless light ray, therefore,  can ever be extracted from a 
black hole. When it comes to rotating black holes such as the Kerr
family of solutions, however, things are not so simple and indeed energy 
can be extracted from black holes as was first noted by Penrose [6].  
Briefly, this energy extraction mechanism proposed by Penrose and 
hence is called ``Penrose process" [6] can be understood as follows.
In Kerr geometry, the surface on which $g_{tt}$ vanishes does not
coincide with the event horizon except at the poles. The toroidal space
inbetween the two surfaces is called ``ergosphere" and in particular the
outer boundary of this ergosphere on which $g_{tt}$ vanishes is dubbed
``static limit" because it can be seen that inside of which no observer 
can possibly remain static. Namely the time translational Killing field 
$\xi^{\mu} = (\partial/\partial t)^{\mu}$ 
becomes spacelike inside the ergosphere
and so does the conserved component $p_{t}$ of the four momentum.
As a consequence, the energy of a particle in this ergoregion, as 
perceived by an observer at infinity, can  be negative. This last
fact leads to a peculiar possibility that, in principle, one can
devise a physical process which extracts energy and angular momentum from
the black hole. The Penrose process, however, requires a precisely
timed breakup of the incident particle at the relativistic velocities
and thus is not a very practical energy extraction scheme.
Perhaps because of this reason, an alternative study of energy
extraction mechanism, known as ``superradiant scattering" [7]
was considered.
In a sense, it can be thought of as a wave analogue of the Penrose process.
If a wave is incident upon a black hole, the part of the wave (``transmitted"
wave) will be absorbed by the black hole and the part of the wave
(``reflected" wave) will escape back to infinity. Normally, the transmitted
wave will carry positive energy into the black hole and the reflected wave
will have less energy than the incident wave. However, for a scalar wave
with the time ($t$) and azimuthal angle ($\phi$) dependence given by
$e^{i(m\phi - \omega t)}$ (with $m$ and $\omega$ being
the azimuthal number and the frequency respectively), the transmitted
wave will carry negative energy into the black hole and the reflected wave
will escape to infinity with greater amplitude and energy than the
originally incident one provided the scalar wave has the frequency in
the range [10]
\begin{eqnarray}
0 < \omega < m\Omega_{H}  \nonumber
\end{eqnarray}
where again $\Omega_{H}$ denotes the angular velocity of the rotating hole
at the event horizon. The ``scalar waves" such as electromagnetic and
gravitational waves exhibit this superradiance [8] when they have frequency
in the range given above. Curiously enough, it is known that fermion
fields do not display superradiance [9]. \\
The spinning AdS$_3$ black hole spacetime that we are considering 
also possesses
very similar structure to that of the Kerr black hole in 4-dim.
including particularly the existence of ergoregion in which the
time translational Killing field $\xi^{\mu}$ becomes spacelike.
Consequently, the natural question one comes to ask is whether or not
there are superradiant scatterings off this AdS$_3$ black hole.
Moreover, we are in a perfect position to quantitatively check the
possibility of superradiance since the precise mode expansions
of the scalar (both real and complex) and the spinor fields are
available now. In other words, unlike the case of scalar or spinor
field in the Kerr black hole spacetime in 4-dim., now we need not
assume the {\it ad hoc} time and azimuthal angle dependences of
the fields to be $e^{i(m\phi - \omega t)}$ but we can pick a
scalar or spinor wave with definite frequency
to check the occurrence of superradiance. As mentioned earlier
in the introduction and as we shall see shortly, as long as
fields are conformally-coupled, the superradiant scattering off
the AdS$_3$ black hole spacetime is absent. 
Therefore in order to demonstrate this absence of superradiance,
in the following subsection, a succinct superradiance-checking
algorithm employing the particle number or energy current will be
formally reviewed and then applied to our AdS$_3$ black hole case. 
\\
{\rm\bf 2. A simple superradiance-checking algorithm}
\\
In this subsection, we would like to first set up a general formalism [10]
that allows us to determine whether or not the superradiance is
actually present in the case of scalar or fermion field. 
To this end, we introduce two quantities of central importance,
``energy current" and ``particle number current".
First, we begin with the energy current. Generally, the ``energy
current" of a field in curved background spacetime is defined by [10]
\begin{eqnarray}
J_{\mu} \equiv - T_{\mu\nu} \xi^{\nu}
\end{eqnarray}
with $\xi^{\mu} = (\partial/\partial t)^{\mu}$ being the time
translational Killing field of a stationary, axisymmetric 
spacetime (which is the Kerr black hole spacetime
for our case). This quantity is obviously conserved owing to
the energy-momentum conservation and the Killing equation
$\nabla^{\mu}\xi^{\nu} + \nabla^{\nu}\xi^{\mu} = 0$
satisfied by the Killing field $\xi^{\mu}$, i.e.,
\begin{eqnarray}
\nabla^{\mu}J_{\mu} = - (\nabla^{\mu}T_{\mu\nu})\xi^{\nu}
- T_{\mu\nu}(\nabla^{\mu}\xi^{\nu}) = 0. \nonumber
\end{eqnarray}
Next, we turn to the particle number current. Generally
speaking, for field theories with action possessing the
global U(1) transformation (i.e., phase transformation)
symmetry, (e.g. complex scalar field theory and fermion
field theory) the associated Noether current can be identified
with the particle number current. Namely the Noether current
of the typical form
\begin{eqnarray}
j^{\mu} = {\delta {\cal L}\over \delta (\nabla_{\mu}
\phi^{i})} \delta \phi^{i} 
\end{eqnarray}
(where ${\cal L}$ denotes the Lagrangian density) is defined
to be the particle number density. Then this particle number
density is covariantly conserved as well due to the
Euler-Lagrange's equation of motion and the invariance of the
action, $\nabla_{\mu}j^{\mu} = 0$. \\
Now in order eventually to determine the presence or absence of
the superradiant scattering, we consider a region $K$ of
spacetime of which the boundary consists of two spacelike
hypersurfaces $\Sigma_{1}$ at ($t$) and  $\Sigma_{2}$ at 
($t+\delta t$) (the constant time slice  $\Sigma_{2}$
is a time translate of  $\Sigma_{1}$ by $\delta t$) and
two timelike hypersurfaces $H$ (black hole horizon at
$r = r_{+}$) and $S_{\infty}$ (large sphere at spatial infinity 
$r \rightarrow \infty$). The appropriate directions of the 
hypersurface normal vector $n^{\mu}$ on each part of the 
boundary are ; $n^{\mu}$ is future-directed on $\Sigma_{1}$,
past-directed on  $\Sigma_{2}$. It is pointing inward the
black hole on the event horizon $H$ and pointing outward to
infinity on  $S_{\infty}$. Then next, consider integrating
the quantity $\nabla^{\mu}J_{\mu}$ (which leads to the
``energy flux" crossing each part of the boundary upon
utilizing the Gauss's theorem) or  $\nabla_{\mu}j^{\mu}$
(which leads to the ``particle number current") over the
region $K$ of spacetime. By using Gauss's theorem we have
\begin{eqnarray}
0 &=& \int_{K} d^4x \sqrt{g} \nabla_{\mu}j^{\mu} \nonumber \\
&=& \int_{\partial K} d^3x \sqrt{h} n_{\mu}j^{\mu} \\
&=& \int_{\Sigma_{1}(t)}n_{\mu}j^{\mu} + \int_{\Sigma_{2}(t+\delta t)} 
n_{\mu}j^{\mu} + \int_{H(r_{+})} n_{\mu}j^{\mu} + 
\int_{S_{\infty}} n_{\mu}j^{\mu} \nonumber
\end{eqnarray}
where $h_{\mu\nu}$ denotes the 3-metric induced on the boundary
$\partial K$ of the region $K$.
Now the terms in the last line of eq.(89) need some explanations.
For boson or fermion field with time dependence of the form
$e^{-i\omega t}$ (which we shall assume throughout) the first two
terms cancel with each other by time translation symmetry.
The third term represents the net particle number flow into the rotating 
black
hole while the last term stands for the net particle number flow out of $K$
to infinity, i.e., the outgoing minus incoming particle number
through $S_{\infty}$ during the time $\delta t$. Thus we end up with
the result 
\begin{eqnarray}
\int_{S_{\infty}} n_{\mu}j^{\mu} = - \int_{H(r_{+})} n_{\mu}j^{\mu}
\end{eqnarray}
which states that the {\it net particle number flow out of $K$ or the
``outgoing minus incoming particle number" equals ``minus" of the
net particle number flow into the rotating black hole}.
Therefore, now we can establish the criterion for the occurrence of
superradiant scattering ; If the quantity on the right hand side
$\int_{H(r_{+})} n_{\mu}j^{\mu}$, namely the net particle number flowing
down the hole, is negative (zero or positive), it means that the
outgoing particle number flux is greater (smaller) than the incident
one and hence the superradiance is present (absent). \\
Thus far we have established the criterion for the occurrence of
superradiance in terms of the ``particle number current" $j^{\mu}$.
An equivalent criterion can be derived in terms of the ``energy 
current" $J_{\mu}$ if we replace  $n_{\mu}j^{\mu}$ by
$<n^{\mu}J_{\mu}>$ (where $<...>$ denotes time averaged quantity)
and replace ``particle number current" with ``energy current"
respectively in the above formalism. \\
Also note that the hypersurface normal $n^{\mu}$ on the black hole
event horizon $H$ is pointing inward the hole and hence is opposite
to the direction of the Killing field [10]
\begin{eqnarray}
\chi^{\mu} = \xi^{\mu} + \Omega_{H} \psi^{\mu}
\end{eqnarray}
which is outer normal to the rotating hole's event horizon.
Thus our task of checking the presence or absence of superradiance
reduces to the computation of the net particle number (or energy)
current flowing into the rotating hole through its event horizon
\begin{eqnarray}
\int_{H(r_{+})} n_{\mu}j^{\mu} = - \int_{H(r_{+})} \chi_{\mu}j^{\mu}.
\end{eqnarray}
Now, before we demonstrate the absence of the superradiance in the case
of AdS$_3$ black hole spacetime, it will be comparative to
illustrate the presence of the superradiance in a boson field case
in the background of Kerr black hole in 4-dim.
using the superradiance-checking formalism introduced above. \\
Thus consider a complex scalar field theory in a stationary, axisymmetric
background spacetime (which we take to be the Kerr black hole 
geometry) described by the action (remember that here in this work, 
we employ the
Misner-Thorne-Wheeler sign convention [1] in which the metric has the
sign of $g_{\mu\nu} = diag (- + + +)$)
\begin{eqnarray}
S = - \int d^4x \sqrt{g} [\nabla_{\mu}\Phi^{\ast} \nabla^{\mu}\Phi +
(M^2 + \xi R)\Phi^{\ast}\Phi]
\end{eqnarray}
and the classical field equations
\begin{eqnarray}
&\nabla_{\mu}&\nabla^{\mu}\Phi - (M^2 + \xi R)\Phi = 0, \\
&\nabla_{\mu}&\nabla^{\mu}\Phi^{\ast} - (M^2 + \xi R)\Phi^{\ast} = 0 
\nonumber
\end{eqnarray}
where $\xi$, $M$ and $R$ denotes some constant (for example, $\xi =
1/6$ with $M = 0$ corresponds to ``conformal coupling''), 
the mass of the scalar 
field and the scalar curvature of the background spacetime respectively.
Then we consider a situation when a complex scalar wave with particular
frequency
\begin{eqnarray}
\Phi(x) = \Phi_{0}(r,\theta)e^{i(m\phi -\omega t)} 
\end{eqnarray}
is incident on and reflected by the Kerr black hole.
Since the Lagrangian density of this complex scalar field 
in eq.(60) is invariant
under the global U(1) (or phase) transformation
\begin{eqnarray}
\Phi(x) &\rightarrow & e^{-i\alpha}\Phi(x), \nonumber \\
\Phi^{\ast}(x) &\rightarrow & e^{i\alpha}\Phi^{\ast}(x) \nonumber
\end{eqnarray}
corresponding Noether current exists and it is
\begin{eqnarray}
j^{\mu} = -i (\Phi^{\ast} \nabla^{\mu}\Phi - \Phi \nabla^{\mu}\Phi^{\ast}).
\end{eqnarray}
This Noether current is the particle number current and it can be seen to
be conserved owing to the classical field equations in eq.(61)
\begin{eqnarray}
\nabla_{\mu}j^{\mu} = 0. \nonumber
\end{eqnarray}
According to the criterion for the occurrence of the superradiance stated
earlier, all we have to do now is to evaluate the net particle number
flowing into the black hole, $\int_{H(r_{+})}n_{\mu}j^{\mu}$ and see if it 
can be negative. Thus on the horizon $r = r_{+}$, we compute the particle
number flux and it is
\begin{eqnarray}
n_{\mu}j^{\mu} &=& - \chi^{\mu}j_{\mu} \nonumber \\
&=& i (\Phi^{\ast}\chi^{\mu}\nabla_{\mu}\Phi - 
\Phi \chi^{\mu}\nabla_{\mu} \Phi^{\ast}) \\
&=& i [\Phi^{\ast}({\partial \over \partial t} + \Omega_{H}
{\partial \over \partial \phi})\Phi -
\Phi ({\partial \over \partial t} + \Omega_{H}
{\partial \over \partial \phi})\Phi^{\ast}] \nonumber \\
&=& 2 (\omega - m\Omega_{H})\mid \Phi_{0} \mid^{2}. \nonumber
\end{eqnarray}
Thus for a complex scalar field with frequency in the range
\begin{eqnarray}
0 < \omega < m\Omega_{H}
\end{eqnarray}
the net particle number flowing down the hole is negative and hence
\begin{eqnarray}
\int_{S_{\infty}} n_{\mu}j^{\mu} =
- \int_{H(r_{+})}  n_{\mu}j^{\mu} > 0
\end{eqnarray}
namely the outgoing minus incident particle number flux through the
large sphere $S_{\infty}$ is positive indicating the occurrence
of superradiance in the case of a scalar field.
\\
{\rm\bf 3. Absence of superradiant scattering off AdS$_3$ black hole}
\\
As usual, we would like to check the occurrence of superradiance
with both boson and fermion fields. Thus we shall consider the
three cases when the real, complex scalar and spinor fields are
scattered off the rotating AdS$_3$ black hole respectively. 
There is, however, a crucial point that we would like to stress 
again.
In 4-dim., we {\it assumed} that both the boson and the fermion
fields possess the time and azimuthal angle dependences given by
$e^{i(m\phi - \omega t)}$. This choice of the form of scalar
and spinor waves is by no means rigorous in that we never really
confirmed that solutions to the Klein-Gordon equation for the
scalar field and solutions to the Dirac equation for the spinor
field in the background of Kerr black hole do take this time
and azimuthal angle dependences. We simply assumed this type
of dependence based on the fact that Kerr black hole spacetime
has time-translational and rotational isometries generated by
Killing fields $\xi^{\mu} = (\partial/\partial t)^{\mu}$ and
$\psi^{\mu} = (\partial/\partial \phi)^{\mu}$ respectively.
For the present case of rotating AdS$_3$ black hole, however,
as we have seen in the preceding section, exact mode expansions
for both scalar and spinor fields are available. Namely, we
do know the precise solution forms of the Klein-Gordon and
Dirac equations in the background of the rotating AdS$_3$
black hole spacetime. Therefore we shall naturally use these
exact solutions as test scalar and spinor waves to be scattered
off the rotating AdS$_3$ black hole. We believe that this
use of actual solutions will provide the correct answer to
our question on the occurrence of superradiance in the case
of AdS$_3$ black hole. \\
First we start with the case with real scalar field.
As we have seen, the mode expansion of real scalar field coupled
conformally to the background of spinning AdS$_3$ black hole is
given in eq.(45).
Since this mode expansion is the superposition of plane waves, we
pick one plane wave solution with particular frequency
\begin{eqnarray}
\Phi(z) = {\rm Re}[({l\over z_2})^{-1/2}e^{ik_{\mu}z^{\mu}}]
= ({l\over z_2})^{-1/2} \cos (k_{\mu}z^{\mu})
\end{eqnarray}
and consider its scattering off the spinning AdS$_3$ black hole.
In order to check the existence of superradiant scattering, we need 
to evaluate the energy flux flowing into the black hole across the
horizon and see if it can be negative under certain circumstances.
Thus on the horizon, $r = r_{+}$, using
\begin{eqnarray}
\Phi(t,r_{+},\phi) = \sum_{n=-\infty}^{\infty} l^{-1/2} e^{{1\over 2}
(la_{+}\phi_{n} - a_{-}t)}
\cos [k_2 e^{(la_{+}\phi_{n} - a_{-}t)}] \nonumber
\end{eqnarray}
the time-averaged energy flux is 
\begin{eqnarray}
<n^{\mu}J_{\mu}> &=& - <\chi^{\mu}J_{\mu}> \nonumber \\
&=& <T_{\mu\nu}\chi^{\mu}\xi^{\nu}> = <(\chi^{\mu}\nabla_{\mu}\Phi)
(\xi^{\nu}\nabla_{\nu}\Phi)> = <(\partial_{t} + \Omega_{H}\partial_{\phi})
\Phi \partial_{t}\Phi> \\
&=& a_{-}(a_{-} - \Omega_{H}la_{+}) 
<\sum_{n=-\infty}^{\infty}\sum_{m=-\infty}^{\infty}
A(t, r_{+}, \phi_{n})A(t, r_{+}, \phi_{m}) > \nonumber
\end{eqnarray}
where
\begin{eqnarray}
A(t, r_{+}, \phi_{n})=l^{-1/2}e^{{1\over 2}(la_{+}\phi_{n} - a_{-}t)}\{{1\over 2}
\cos [k_2 e^{(la_{+}\phi_{n} - a_{-}t)}] - k_{2}e^{(la_{+}\phi_{n} - a_{-}t)}
\sin [k_2 e^{(la_{+}\phi_{n} - a_{-}t)}]\}. \nonumber
\end{eqnarray}
Now, if the quantity in the last line could be negative, then the 
superradiance occurs. Obviously, however, this cannot really happen since
$\Omega_{H} = a/r_{+}^2 = r_{-}/lr_{+}$ and $a_{\pm} = r_{\pm}/l^2$,
\begin{eqnarray}
(a_{-} - \Omega_{H} l a_{+}) = 0.
\end{eqnarray}
Therefore, regardless of the value of its frequency, (real) scalar field 
does not display superradiance. And this is in contrast to what happens
in the case of Kerr black holes in 4-dim. spacetime. \\
Secondly, we consider the case of complex scalar field.
The mode expansion of complex scalar field coupled
conformally to the background of spinning AdS$_3$ black hole is
given in eq.(46).
Again, since these mode expansions are the superpositions of plane waves, we
pick one plane wave solution with particular frequency
\begin{eqnarray}
\Phi(z) &=& ({l\over z_2})^{-1/2}e^{ik_{\mu}z^{\mu}}, \\ 
\Phi^{\ast}(z) &=& ({l\over z_2})^{-1/2}e^{-ik_{\mu}z^{\mu}}. \nonumber
\end{eqnarray}
and consider its scattering off the spinning AdS$_3$ black hole.
In order to see if there is a  superradiant scattering, we need
to compute the particle number flux flowing into the black hole across the
horizon and check whether it can be negative under certain circumstances.
Thus on the horizon, $r = r_{+}$, using
\begin{eqnarray}
\Phi(t,r_{+},\phi) &=& \sum_{n=-\infty}^{\infty}l^{-1/2} 
e^{{1\over 2}(la_{+}\phi_{n} - a_{-}t)}
\exp [ik_2 e^{(la_{+}\phi_{n} - a_{-}t)}], \nonumber \\
\Phi^{\ast}(t,r_{+},\phi) &=& \sum_{n=-\infty}^{\infty}l^{-1/2} 
e^{{1\over 2}(la_{+}\phi_{n} - a_{-}t)}
\exp [-ik_2 e^{(la_{+}\phi_{n} - a_{-}t)}]. \nonumber
\end{eqnarray}
the particle number flux is
\begin{eqnarray}
n_{\mu}j^{\mu} &=& - \chi^{\mu}j_{\mu} \nonumber \\
&=& i(\Phi^{\ast}\chi^{\mu}\nabla_{\mu}\Phi -
\Phi \chi^{\mu}\nabla_{\mu}\Phi^{\ast}) \\
&=& i[\Phi^{\ast}(\partial_{t} + \Omega_{H}\partial_{\phi})\Phi -
\Phi(\partial_{t} + \Omega_{H}\partial_{\phi})\Phi^{\ast}] \nonumber \\
&=& (a_{-} - \Omega_{H}la_{+})k_{2}\sum_{n=-\infty}^{\infty}
\sum_{m=-\infty}^{\infty}B(t, r_{+}, \phi_{n})B^{\ast}(t, r_{+}, \phi_{m})
[e^{(la_{+}\phi_{n} - a_{-}t)} + e^{(la_{+}\phi_{m} - a_{-}t)}] \nonumber
\end{eqnarray}
where 
$B(t, r_{+}, \phi_{n})=l^{-1/2}e^{{1\over 2}(la_{+}\phi_{n} - a_{-}t)}
\exp [ik_2 e^{(la_{+}\phi_{n} - a_{-}t)}]$.
Again, if the quantity in the last line could be negative, then the
superradiance occurs which, as we have realized earlier in the case of
real scalar field, cannot really happen since
$\Omega_{H} = a/r_{+}^2 = r_{-}/lr_{+}$ and $a_{\pm} = r_{\pm}/l^2$, thus
\begin{eqnarray}
(a_{-} - \Omega_{H} l a_{+}) = 0.
\end{eqnarray}
Again, irrespective of the value of its frequency, complex scalar field
does not display superradiance either. \\
Finally, we turn to the case of spinor field.
As already mentioned and as we shall see shortly as well, in
order to check the occurrence of superradiance in the 
fermion field case, one needs the concrete geometry structure
of the background AdS$_3$ black hole. Besides, the standard
formulation of spinor field theory in curved background spacetime
is associated with the Riemann-Cartan formulation 
of general relativity in which one of the basic
computational tools is the use of the non-holonomic basis 1-form
(i.e., ``soldering form"). Thus here we begin with the
AdS$_3$ black hole metric (given in Schwarzschild-like coordinates)
written in the ADM's (2+1) space-plus-time split form which proves to be
suitable to be converted to the one in non-coordinate basis
\begin{eqnarray}
ds^2 &=& -N^2 dt^2 + h_{rr}dr^2 +
h_{\phi \phi}[d\phi + N^{\phi}dt]^2 \nonumber \\
&=& g_{\mu\nu}dx^{\mu}dx^{\nu} = \eta_{ab}e^{a}e^{b} 
\end{eqnarray}
where Greek indices refer to the accelerated frame of reference 
(i.e., coordinate basis, $\mu = t,~r,~\phi$) and
the Roman indices refer to the locally inertial reference frame
(i.e., non-coordinate basis, $a = 0,~1,~2$). 
Also we used the definitions for
the soldering form (``dreibein") $g_{\mu\nu} = \eta_{ab}
e^{a}_{\mu}e^{b}_{\nu}$ and the non-coordinate basis 1-form
$e^{a} = e^{a}_{\mu}dx^{\mu}$. In the ADM's (2+1) split form
above, the lapse, shift functions and the spatial metric components
are given respectively by
\begin{eqnarray}
N^2(r) &=& (-M + {r^2 \over l^2} + {J^2 \over 4r^2}), \nonumber \\
N^{\phi}(r) &=& -{J \over 2r^2}, ~~~N^{r}(r) = 0, \nonumber \\
h_{rr}(r) &=& N^{-2}(r), ~~~h_{\phi \phi}(r) = r^2, \\ 
h_{r\phi} &=& h_{\phi r} = 0. \nonumber
\end{eqnarray}
where $M$ and $J = 2a$ denote the ADM mass and the angular momentum
per unit mass of the AdS$_3$ black hole respectively and 
$l^{-2} = (-\Lambda)$ as indicated earlier.  As is well known the
event horizon developes at the larger zero of $N^2(r_{+}) = 0$.\\
Actually, the virtue of writing the AdS$_3$ black hole metric in the ADM's
(2+1) split form as in eq.(73) above is that from which now one can 
read off the non-coordinate basis 1-form easily as follows
\begin{eqnarray}
e^{0} &=& e^{0}_{\mu}dx^{\mu} = N dt, \nonumber \\
e^{1} &=& e^{1}_{\mu}dx^{\mu} = \sqrt{h_{rr}} dr = N^{-1} dr, \\
e^{2} &=& e^{2}_{\mu}dx^{\mu} = \sqrt{h_{\phi \phi}}(d\phi + N^{\phi}dt)
= r(d\phi + N^{\phi}dt). \nonumber
\end{eqnarray}
Equvalently, the dreibein and the inverse dreibein can be read off as
\begin{equation}
e^{a}_{\mu} = \left( \begin{array}{ccc}
                              N & 0 & 0 \\
                              0 & N^{-1} & 0 \\
                              rN^{\phi} & 0 & r \\ 
                     \end{array} \right),
~~~e^{\mu}_{a} = \left( \begin{array}{ccc}
                              N^{-1} & 0 & 0 \\
                              0 & N & 0 \\
                              -N^{-1}N^{\phi} & 0 & r^{-1} \\
                     \end{array} \right)
\end{equation}
Further, the spin connection 1-form $\omega^{ab} = \omega^{ab}_{\mu}
dx^{\mu}$ can be obtained from the Cartan's 1st structure equation,
$de^{a} + \omega^{a}_{b}\wedge e^{b} = 0$ using the non-coodinate
basis 1-form given in eq.(75).  Here, however, we do not
look for the spin connection since we shall not really need its
explicit form in the discussion below leading to the conclusion on
the absence of fermionic superradiance. \\
Now, for later use we write the $\gamma$-matrices with coordinate basis
indices in accelerated frame (i.e., in Schwarzschild-type coordinates)
in terms of those with non-coordinate basis indices in locally-inertial
frame using the soldering form (inverse dreibein) given above, i.e.,
$\gamma^{\mu}(x) = e^{\mu}_{a}(x)\gamma^{a}$
\begin{eqnarray}
\gamma^{t} &=&  e^{t}_{a}\gamma^{a} = N^{-1}\gamma^{0}, \nonumber \\
\gamma^{r} &=&  e^{r}_{a}\gamma^{a} = N\gamma^{1}, \\
\gamma^{\phi} &=&  e^{\phi}_{a}\gamma^{a} = -N^{-1}N^{\phi}\gamma^{0}
+ r^{-1}\gamma^{2}. \nonumber
\end{eqnarray}
With this preparation, now we consider the conformally-coupled 
spinor field theory in
the background of this AdS$_3$ black hole spacetime described 
by the action [2]
\begin{eqnarray}
S = \int d^3x \sqrt{g} \{ {i\over 2}[\bar{\psi}\gamma^{\mu}
\overrightarrow{\nabla}_{\mu}\psi - \bar{\psi}\gamma^{\mu}
\overleftarrow{\nabla}_{\mu}\psi]\}
\end{eqnarray}
and the classical field equations, i.e., curved spacetime Dirac equations
\begin{eqnarray}
i\gamma^{\mu}\overrightarrow{\nabla}_{\mu} \psi = 0, 
~~~\bar{\psi} i\gamma^{\nu}\overleftarrow{\nabla}_{\nu}  = 0.  
\end{eqnarray}
The mode expansion of spinor field coupled conformally to the 
background of spinning AdS$_3$ black hole was given in eq.(47).
Again, since this mode expansion is the superposition of
plane waves,
we consider a situation when a spinor wave with particular
frequency and spin
\begin{eqnarray}
\psi(z) = ({l\over z_{2}})^{-1}u(p)e^{ip_{\mu}z^{\mu}}
\end{eqnarray}
is incident on and reflected by the AdS$_3$ black hole.
Here the 2-component
Dirac spinor $u(p)$ satisfies the Dirac equation in curved
spacetime given above. As usual, in order to check the
occurrence of superradiant scattering, we need to evaluate
the particle number flux across the event horizon and see
if it can be negative under certain circumstances.
To this end, we begin by defining the particle number current
for spinor field.
Since the Lagrangian density of this spinor field 
given in eq.(78) is also 
invariant under the global U(1) (or phase) transformation
\begin{eqnarray}
\psi(x) \rightarrow  e^{-i\alpha}\psi(x), 
~~~\bar{\psi}(x) \rightarrow  \bar{\psi}(x)e^{i\alpha} \nonumber
\end{eqnarray}
corresponding Noether current exists and it is
\begin{eqnarray}
j^{\mu} = {\delta {\cal L}\over 
\delta (\overrightarrow{\nabla}_{\mu}\psi)}\delta \psi
+ \delta \bar{\psi} 
{\delta {\cal L}\over 
\delta (\bar{\psi}\overleftarrow{\nabla}_{\mu})} 
= \bar{\psi} \gamma^{\mu} \psi. 
\end{eqnarray}
This Noether current is identified with the particle number current
and it can be seen to be conserved due to the Dirac equations given above
\begin{eqnarray}
\nabla_{\mu}j^{\mu} = \bar{\psi}\gamma^{\mu}\overleftarrow{\nabla}_{\mu}
\psi + \bar{\psi}\gamma^{\mu}\overrightarrow{\nabla}_{\mu}\psi = 0 \nonumber
\end{eqnarray}
Thus on the horizon, $r = r_{+}$, using
\begin{eqnarray}
\psi(t,r_{+},\phi) &=& \sum_{n=-\infty}^{\infty}l^{-1} 
e^{(la_{+}\phi_{n} - a_{-}t)}u(p)
\exp [ip_2 e^{(la_{+}\phi_{n} - a_{-}t)}], \nonumber \\
\bar{\psi}(t,r_{+},\phi) &=& \sum_{n=-\infty}^{\infty}l^{-1} 
e^{(la_{+}\phi_{n} - a_{-}t)}\bar{u}(p)
\exp [-ip_2 e^{(la_{+}\phi_{n} - a_{-}t)}] \nonumber
\end{eqnarray}
the particle number flux is
\begin{eqnarray}
n_{\mu}j^{\mu} &=& - \chi^{\mu}j_{\mu} = - \bar{\psi} g_{\alpha\beta}
\chi^{\alpha}\gamma^{\beta}\psi \nonumber \\
&=&  - \bar{\psi} g_{\alpha\beta} (\delta^{\alpha}_{t} + 
\Omega_{H}\delta^{\alpha}_{\phi}) \gamma^{\beta} \psi \\
&=& - \bar{\psi}[(g_{tt} + \Omega_{H}g_{t\phi})\gamma^{t} +
(g_{t\phi} + \Omega_{H}g_{\phi\phi})\gamma^{\phi}]\psi \nonumber \\
&=& - \bar{\psi}[ - N\gamma^{0} + r(N^{\phi} + \Omega_{H})\gamma^{2}]
\psi = 0. \nonumber
\end{eqnarray}
where we used the relation between $\gamma$-matrices with coordinate basis
indices in accelerated frame (i.e., in Schwarzschild-type coordinates)
and those with non-coordinate basis indices in locally-inertial
frame, $\gamma^{\mu}(x) = e^{\mu}_{a}(x)\gamma^{a}$ derived earlier
in eq.(77) and $g_{tt} = -[N^2-r^2(N^{\phi})^2]$, $g_{t\phi} = r^2N^{\phi}$
and $g_{\phi\phi} = r^2$. And to get the last equality to zero we used
\begin{eqnarray}
N^2(r_{+}) = 0, ~~~N^{\phi}(r_{+}) = -({a\over r^2_{+}}) = 
- \Omega_{H}. \nonumber
\end{eqnarray}
This result indicates that the net fermionic particle number flux flowing 
down the hole through its event horizon is zero {\it irrespective of
the frequency of the fermion field}.
Therefore the outgoing minus incident fermionic particle
number flux through the large sphere $S_{\infty}$ is zero
\begin{eqnarray}
\int_{S_{\infty}} n_{\mu}j^{\mu} = - \int_{H(r_{+})}n_{\mu}j^{\mu} = 0
\nonumber
\end{eqnarray}
establishing again the absence of superradiance in the case of 
fermion field as well. \\
Thus far we have illustrated the absence of fermionic superradiance in 
terms of the particle number flux. One can draw the same conclusion in
terms of the energy flux we introduced earlier by showing that it also
is zero through the event horizon of the AdS$_3$ black hole.
Since it is rather straightforward to demonstrate this, we shall not
get into the detail further. \\
We now wish to provide physical interpretation of the absence of both
bosonic and fermionic superradiances observed above. To this end, we
should go back and cite our earlier statement on 
the physical meaning of the conformal vacuum. 
In curved spacetimes, generally there is no meaningful notion of global
vacuum and global Fock space. The vacuum and hence the concept of particle
is really observer-dependent. Nevertheless, if there exist geometrical
symmetries in the background spacetime, it may be that a particular set
of modes and the corresponding vacuum and Fock space emerge as having
natural physical meaning. The conformally-coupled scalar and spinor
fields in conformally-flat spacetimes like AdS$_3$ black hole spacetime
are endowed with such a feature and the associated vacuum state is the
``conformal vacuum'' which remains to be a vacuum with respect to
any other reference frame. This means that particle production is 
absent since $\mid 0, {\rm in}>$ and $\mid 0, {\rm out}>$ are identical.
In view of this, the absence of the superradiance which is a stimulated
emission phenomenon from the ergoregion seems to be a natural 
consequence. Furthermore, this observation implies that if one
sticks to consider solely the conformally-coupled fields in this AdS$_3$ black
hole spacetime, the Hawking radiation [13], namely the spontaneous emission 
of particles should be absent as well. 

\vspace*{0.5cm}

\noindent
\begin{center}
{\rm\bf VI. Discussions}
\end{center}

Now we summarize what has been done in this work. 
Noticing that the AdS$_3$ black hole solution
discovered recently by BTZ [3] has attracted a lot of interest currently,
we attempted quantisation of scalar and spinor fields in the background
of this AdS$_3$ black hole spacetime. And to do so we particularly employed
the ``conformal gauge'' in which the metric takes the manifestly
conformally-flat form. Mode expansions (given in sect.III) 
and two point Green's functions (which will be given in the appendix below)
for the scalar and spinor fields have been obtained in closed forms.
To be more concrete, first the construction of two-point Green's
functions for conformally-coupled fields performed in the present
work can be thought of as employing the ``transparent boundary
conditions'' introduced originally by Avis, Isham and Storey [12]. 
Besides the construction of the Green's functions for conformally-coupled
spinor field given in the present work has not yet been attempted in the
literature. It {\it is} a new contribution. Next, the
construction of mode expansions for quantum fields in explicit, closed
forms allowed several important observations. Firstly, the particle 
number conservation across the event horizon and the Cauchy horizon 
of the AdS$_3$ black hole
confirms that indeed the inner and outer horizons are just coordinate 
singularities on which nothing special happens. Secondly, but
more importantly, on the absence of superradiant scattering off the
spinning AdS$_3$ black hole. Since the spinning AdS$_3$ black hole
possesses very similar causal structure to that of Kerr black hole
in 4-dim. including the existence of ergoregion, one naturally
gets interested in the possibility of superradiance phenomenon.
However, unlike the conventional study of superradiance for the Kerr 
black hole in 4-dim., now one needs not assume the {\it ad hoc}
time and azimuthal angle dependences of the fields to be
$e^{i(m\phi - \omega t)}$ since the precise mode expansion forms for
the scalar and spinor fields are known. Thus using these exact 
solutions to the Klein-Gordon and Dirac equations in the background
of the AdS$_3$ black hole spacetime as test scalar and spinor waves to 
be scattered off the rotating AdS$_3$ black hole, both the bosonic
and fermionic superradiances have been displayed to be absent.
The physical interpretation of this absence of superradiance
(i.e., the stimulated emission) and the absence of Hawking evaporation
(i.e., the spontaneous emission) of AdS$_3$ black hole for the case
of ``conformal triviality'' (namely, the case of conformally-coupled
fields in conformally-flat spacetime), then, has been provided in
terms of the conformal vacuum, which is unique and has global meaning.
Finally, it is our hope that our general formulation of the quantisation 
of conformally-coupled fields in conformally flat spacetimes in arbitrary
dimensions and our explicit demonstration particularly in AdS$_3$ black
hole spacetime may find interesting applications in various other situations.

\begin{center}
{\large\bf Acknowledgements}
\end{center}
H.K. was supported in part by Korea Research Foundation and Basic Science
Research Institute at Ewha Women's Univ (BSRI - 97 - 2427) and at 
Sogang Univ (BSRI-96-2414). 

\newpage

\begin{center}
{\rm\bf Appendix : The Green's functions of conformally-coupled  fields }
\end{center}

Here, we shall provide the 2-point Green's functions, particularly
the positive-frequency Wightman functions [2] of the scalar and spinor fields
coupled conformally to the spinning AdS$_3$ black hole spacetime in 
closed forms.
\noindent
\\
{\rm\bf 1. Scalar field}
\\
For any 2-point Green's function of scalar field conformally coupled to 
a conformally flat spacetime is given by  (in $\it n$-dimension) [2]

\begin{eqnarray}
 G(x, x') = \Omega^{-({{n-2}\over 2})}(x)
           G_{0}(x, x') 
           \Omega^{-({{n-2}\over 2})}(x')                 
\end{eqnarray}
with $G_0$ being the free field propagator in flat, Minkowski spacetime. 
Consider that we are interested in the positive-frequency Wightman function
for the real scalar field in the background of AdS$_3$ black hole spacetime.
Then it is given by

\begin{eqnarray}
 G^{+}(z,z') = \Omega^{-{1\over 2}}(z)G^{+}_{0}(z,z') 
               \Omega^{-{1\over 2}}(z')                     
\end{eqnarray}
We first have to evaluate, in flat 3-dimensional spacetime  
    
\begin{eqnarray}
 G^{+}_{0}(z,z') = \int {d^{3}k\over (2\pi )^3}
               {-1\over k^2}
               e^{ik\cdot (z-z')}                    
\end{eqnarray}
where a particularly prescribed integration contour in the complex
$k^0$ plane is assumed. However, if we analytically continue to 
the Euclidean space via the ``Wick rotation''
\begin{eqnarray}
       t   \rightarrow t_{E}   = it,      
      ~~~ k^0 \rightarrow k^0_{E} = ik^0         
\end{eqnarray}
then  the evaluation of the integral can be done rather easily to yield

\begin{eqnarray}
 G^{+}_{0}(z,z') &=& i \int {d^{3}k_{E}\over (2\pi )^3}
                  {1\over k^2_{E}}
                  e^{ik^{E}_{\mu}(z-z')^{\mu}}   \\
              &=& {i\over {4\pi}}
                  {1\over {|z-z'|}}           
               = {({i\over 4\pi})}
                  \Bigl[ - (z_0 - {z'}_0)^2 
                         + (z_1 - {z'}_1)^2
                         + (z_2 - {z'}_2)^2 
                  \Bigr]^{-{1\over 2}}.    \nonumber
\end{eqnarray}
Therefore 
 
\begin{eqnarray}
 G^{+}(z,z') &=& \Omega^{-{1\over 2}}(z)
                 G^{+}_{0}(z,z')
                 \Omega^{-{1\over 2}}(z')  
                                                      \nonumber  \\
             &=& {i\over {4\pi l}}
                 \Bigl[ {{ - (z_0 - {z'}_0)^2 
                           + (z_1 - {z'}_1)^2
                           + (z_2 - {z'}_2)^2 }
                         \over 
                         { z_2{z'}_2}}
                 \Bigr]^{-{1\over 2}}. 
\end{eqnarray}
As was stated when obtaining the correct mode expansions of the scalar and
spinor fields conformally coupled to this BTZ black hole spacetime, the correct
Green functions exhibiting the periodicity in $\phi$-coordinate can be
constructed via the infinite linear sum
\begin{eqnarray}
G^{+}(z, z') = \sum_{n=-\infty}^{\infty} G^{+}(z_{n}, z'_{n}) \nonumber
\end{eqnarray}
(where $z_{n}=z(t, r, \phi_{n})$, $z'_{n}=z(t, r, \phi'_{n})$ with again
$\phi_{n}=\phi + 2\pi n$ and $\phi'_{n}=\phi' + 2\pi n$) which amounts to
employing the method of images [15]. \\
(1) For $z,z'\in $ Region (I); $r,r'\ge r_{+}$

\begin{eqnarray}
 G^{+}(z_{n}, z'_{n}) &=& {i\over 4{\sqrt 2}\pi l}
       \Biggl\{ 
              {{ ( r^2 -r^2_{-})^{1\over 2}
                 ( {r'}^2 -r^2_{-})^{1\over 2}}
                \over 
                 {r^2_{+} - r^2_{-} }} \cosh F_{-}(\Delta t, \Delta \phi_{n})
                                                    \nonumber      \\
         && - {{ ( r^2 -r^2_{+})^{1\over 2}
                 ({r'}^2 -r^2_{+})^{1\over 2}}     
                \over 
                 {r^2_{+} - r^2_{-}}} \cosh F_{+}(\Delta t, \Delta \phi_{n})
            - 1 
       \Biggr\}^{-{1\over 2}}.                      
                                                   \nonumber   
\end{eqnarray}
(2) For $z,z'\in $ Region (II); $r_{-}< r,r' \le r_{+}$

\begin{eqnarray}
 G^{+}(z_{n}, z'_{n}) &=& {i\over 4{\sqrt 2}\pi l}
       \Biggl\{ 
              {{ (r^2 -r^2_{-})^{1\over 2}
                 ({r'}^2 -r^2_{-})^{1\over 2}}
                \over 
                 {r^2_{+} - r^2_{-} }} \cosh F_{-}(\Delta t, \Delta \phi_{n})
                                                    \nonumber      \\
         && + {{ (r^2_{+} -r^2)^{1\over 2}
                 (r^2_{+} -{r'}^2)^{1\over 2}}    
                \over 
                 {r^2_{+} - r^2_{-} }} \cosh F_{+}(\Delta t, \Delta \phi_{n})
            - 1 
       \Biggr\}^{-{1\over 2}}.                 
                                                    \nonumber   
\end{eqnarray}
(3) For $z,z'\in $ Region (III); ${\rm 0} < r,r' \le  r_{-}$

\begin{eqnarray}
 G^{+}(z_{n}, z'_{n}) &=& {i\over 4{\sqrt 2}\pi l}
       \Biggl\{ 
              {{ (r^2_{+} - r^2)^{1\over 2}
                 (r^2_{+} - {r'}^2)^{1\over 2}}
                \over 
                 {r^2_{+} - r^2_{-} }} \cosh F_{+}(\Delta t, \Delta \phi_{n})
                                                    \nonumber      \\
         && - {{ (r^2_{-} - r^2)^{1\over 2}
                 (r^2_{-} - {r'}^2)^{1\over 2}}    
                \over 
                 {r^2_{+} - r^2_{-}}} \cosh F_{-}(\Delta t, \Delta \phi_{n})
            - 1 
       \Biggr\}^{-{1\over 2}}.            
                                                    \nonumber   
\end{eqnarray}
(4) For $ z\in $ Region(I); $(r \ge r_{+})\ \ \ $ 
        $ z'\in $ Region (II); $(r_{-} <  r' \le r_{+})$  

\begin{eqnarray}
 G^{+}(z_{n}, z'_{n}) &=& {i\over 4{\sqrt 2}\pi l}
       \Biggl\{ 
              {{ (r^2 - r^2_{-})^{1\over 2}
                 ({r'}^2 - r^2_{-})^{1\over 2}}
                \over 
                 {r^2_{+} - r^2_{-}}} \cosh F_{-}(\Delta t, \Delta \phi_{n})
                                                    \nonumber      \\
         && + {{ (r^2 - r^2_{+})^{1\over 2}
                 (r^2_{+} -{r'}^2)^{1\over 2}}    
                \over 
                 {r^2_{+} - r^2_{-}}} \sinh  F_{+}(\Delta t, \Delta \phi_{n})
            - 1 
       \Biggr\}^{-{1\over 2}}.        
                                                   \nonumber   
\end{eqnarray}
(5) For  $  z\in $ Region (II);  $(r_{-} < r \le r_{+})\ \ \ $ 
         $ z'\in $ Region (III); $({\rm 0} < r' \le r_{-})$ 

\begin{eqnarray}
 G^{+}(z_{n}, z'_{n}) &=& {i\over 4{\sqrt 2}\pi l}
       \Biggl\{ 
                {{ (r^2 - r^2_{-})^{1\over 2}
                      (r^2_{+} - {r'}^2)^{1\over 2}}
                     \over 
                      {r^2_{+} - r^2_{-}}} 
               \cosh [(a_{-}t-a_{+}t') - l(a_{+}\phi_{n} - a_{-}\phi'_{n})]
                                                    \nonumber      \\
              &&- {{ (r^2_{+} - r^2)^{1\over 2}
                     (r^2_{-} -{r'}^2)^{1\over 2}}    
                    \over 
                     {r^2_{+} - r^2_{-} }} 
              \sinh [(a_{+}t-a_{-}t') - l(a_{-}\phi_{n} - a_{+}\phi'_{n})] 
                - 1 
       \Biggr\}^{-{1\over 2}}.    
                                                    \nonumber   
\end{eqnarray}
where we used the short-hand notation 
$ F_{\pm}(\Delta t, \Delta \phi_{n}) \equiv [a_{\pm}(t-t') - 
la_{\mp}(\phi -\phi' + 2\pi n)]$ with $a_{\pm} = r_{\pm}/l^2$ 
and $\phi_{n} = \phi + 2\pi n$, $\Delta \phi_{n} = (\phi - \phi')+2\pi n$ 
as defined earlier.

\vspace*{0.5cm}

\noindent
{\rm\bf 2. Spinor field }
\\
Generally, any 2-point Green's function of spinor field conformally-coupled 
to a conformally-flat spacetime is given by [2]

\begin{eqnarray}
 S(x, x') = \Omega^{-({{n-1}\over 2})}(x)
           S_{0}(x, x') 
           \Omega^{-({{n-1}\over 2})}(x')        
\end{eqnarray}
with $S_0$ being the free, massless field Green's function in flat, 
Minkowski spacetime as mentioned earlier.
Besides, as pointed out earlier, the spinor field Green's 
function in flat spacetime is related to that of the scalar field by
(generally in the presence of the mass)
\begin{eqnarray}
 S_{0}(x, x') = (i\gamma^{\alpha}\partial^{x}_{\alpha} + m )G_{0}(x, x').
\end{eqnarray}
Again, consider that we are interested in the positive-frequency Wightman
function for the spinor field in the background of the AdS$_3$ black hole
spacetime. Then it is given by

\begin{eqnarray}
 S^{+}(z,z') = \Omega^{-1}(z)S^{+}_{0}(z,z')\Omega^{-1}(z') 
\end{eqnarray}
where $S^{+}_{0}$ can be obtained from $G^{+}_{0}$ evaluated earlier as
\begin{eqnarray}
 S^{+}_{0}(z,z') &=& (i\gamma^{\alpha}\partial^{z}_{\alpha})
                     G^{+}_{0}(z,z')  \\
             &=& - (\gamma^{\alpha}\partial^{z}_{\alpha})      
                     \Bigl[{1\over {4\pi}}{1\over {|z-z'|}}\Bigr]    
             = ({1\over {4\pi}})
                     { {\gamma^{\alpha}(z-z')_{\alpha}}\over 
                       {|z-z'|^3} }. \nonumber
\end{eqnarray}
Finally, therefore  
\begin{eqnarray}
 S^{+}(z,z') 
       &=& ({l\over z_2})^{-1}
           \Biggl\{ 
                   {({1\over {4\pi}})}
                   { { - \gamma^0(z_0 - {z'}_0) 
                       + \gamma^1(z_1 - {z'}_1) 
                       + \gamma^2(z_2 - {z'}_2)}
                   \over
                     [ - (z_0 - {z'}_0)^2 
                       + (z_1 - {z'}_1)^2 
                       + (z_2 - {z'}_2)^2 ]^{3\over 2} }
           \Biggr\} 
           ({l\over {z'}_2})^{-1}          \nonumber   \\
       &=& {1\over 4\pi l^2}
           \Biggl\{
                   { { - (z_0 - {z'}_0)^2 
                       + (z_1 - {z'}_1)^2 
                       + (z_2 - {z'}_2)^2 }
                   \over 
                     {z_2{z'}_2} }
           \Biggr\}^{-{3\over 2}}     
                                                   \nonumber  \\
       &&  \times (z_2{z'}_2)^{-{1\over 2}}
           [ - \gamma^0(z_0 - {z'}_0) 
             + \gamma^1(z_1 - {z'}_1) 
             + \gamma^2(z_2 - {z'}_2) ].    
\end{eqnarray}
And again the correct Green function is given by
\begin{eqnarray}
S^{+}(z, z') = \sum_{n=-\infty}^{\infty} S^{+}(z_{n}, z'_{n}) \nonumber
\end{eqnarray}
(1)  For $z,z'\in $ Region (I); $r,r'\ge r_{+}$

\begin{eqnarray}
 (z_0 - {z'}_0) 
 &=& \Bigg\{ 
            \Bigl( 
                  {{r^2 - r^2_{+}}\over {r^2 - r^2_{-}}}
            \Bigr)^{1\over 2}
            \sinh F_{+}(t, \phi_{n}) e^{[-F_{-}(t, \phi_{n})]}
          - \Bigl( 
                  {{{r'}^2 - r^2_{+}}\over {{r'}^2 - r^2_{-}} }
            \Bigr)^{1\over 2}
            \sinh F_{+}(t', \phi'_{n}) e^{[-F_{-}(t', \phi'_{n})]}
     \Bigg\},     
                                                   \nonumber \\
 (z_1 - {z'}_1) 
 &=& \Bigg\{ 
            \Bigl( 
                  {{r^2 - r^2_{+}}\over {r^2 - r^2_{-}} }
            \Bigr)^{1\over 2}
            \cosh F_{+}(t, \phi_{n}) e^{[-F_{-}(t, \phi_{n})]}
          - \Bigl( 
                  {{{r'}^2 - r^2_{+}}\over {{r'}^2 - r^2_{-}} }
            \Bigr)^{1\over 2}
            \cosh F_{+}(t', \phi'_{n}) e^{[-F_{-}(t', \phi'_{n})]}
     \Bigg\},                                
                                                   \nonumber \\ 
 (z_2 - {z'}_2) 
 &=& \Bigg\{ 
            \Bigl( 
                  {{r^2_{+} - r^2_{-}}\over {r^2 - r^2_{-}}}
            \Bigr)^{1\over 2}
            e^{[-F_{-}(t, \phi_{n})]}
          - \Bigl( 
                  {{r^2_{+} - r^2_{-}}\over {{r'}^2 - r^2_{-}}}
            \Bigr)^{1\over 2}
            e^{[-F_{-}(t', \phi'_{n})]}
     \Bigg\}.    
                                                   \nonumber
\end{eqnarray}

\begin{eqnarray}
 S^{+}(z_{n}, z'_{n}) 
 &=& {1\over 8\sqrt 2\pi l^2}
  \Biggl\{
          {{(r^2 -r^2_{-})^{1\over 2}
            ({r'}^2 -r^2_{-})^{1\over 2}}
           \over 
            {r^2_{+} - r^2_{-}} }
         \cosh F_{-}(\Delta t, \Delta \phi_{n})
                                                    \nonumber    \\   
 &&     - {{(r^2 -r^2_{+})^{1\over 2}
            ({r'}^2 -r^2_{+})^{1\over 2}}
           \over 
            {r^2_{+} - r^2_{-}} }
         \cosh F_{+}(\Delta t, \Delta \phi_{n})
        - 1
  \Biggr\}^{-{3\over 2}}
                                                    \nonumber    \\
 &&\times 
  {{(r^2 -r^2_{-})^{1\over 4}
    ({r'}^2 -r^2_{-})^{1\over 4}}
   \over 
   {(r^2_{+} - r^2_{-})^{1\over 2}}}
  \exp{{1\over 2}
  \Bigl[ F_{-}(t, \phi_{n}) + F_{-}(t', \phi'_{n}) \Bigr]}
                                                    \nonumber    \\
 &&\times
  \Bigl[ - \gamma^0(z_0 - {z'}_0) 
         + \gamma^1(z_1 - {z'}_1) 
         + \gamma^2(z_2 - {z'}_2) 
  \Bigr].       
                                                    \nonumber
\end{eqnarray}
(2) For $z,z'\in $ Region (II); $r_{-}< r,r' \le r_{+}$

\begin{eqnarray}
 (z_0 - {z'}_0) 
 &=& \Bigg\{ 
            \Bigl( 
                  {{r^2_{+} - r^2}\over {r^2 - r^2_{-}}}
            \Bigr)^{1\over 2}
            \cosh F_{+}(t, \phi_{n}) e^{[-F_{-}(t, \phi_{n})]}
         - \Bigl( 
                  {{r^2_{+} - {r'}^2}\over {{r'}^2 - r^2_{-}} }
            \Bigr)^{1\over 2}
            \cosh F_{+}(t', \phi'_{n}) e^{[-F_{-}(t', \phi'_{n})]}
     \Bigg\},     
                                                   \nonumber \\
 (z_1 - {z'}_1) 
 &=& \Bigg\{ 
            \Bigl( 
                  {{r^2_{+} - r^2}\over {r^2 - r^2_{-}} }
            \Bigr)^{1\over 2}
            \sinh F_{+}(t, \phi_{n}) e^{[-F_{-}(t, \phi_{n})]}
         - \Bigl( 
                  {{r^2_{+} - {r'}^2}\over {{r'}^2 - r^2_{-}} }
            \Bigr)^{1\over 2}
           \sinh F_{+}(t', \phi'_{n}) e^{[-F_{-}(t', \phi'_{n})]}
     \Bigg\},                                
                                                   \nonumber \\ 
 (z_2 - {z'}_2) 
 &=& \Bigg\{ 
            \Bigl( 
                  {{r^2_{+} - r^2_{-}}\over {r^2 - r^2_{-}}}
            \Bigr)^{1\over 2}
            e^{[-F_{-}(t, \phi_{n})]}
          - \Bigl( 
                  {{r^2_{+} - r^2_{-}}\over {{r'}^2 - r^2_{-}}}
            \Bigr)^{1\over 2}
             e^{[-F_{-}(t', \phi'_{n})]}
     \Bigg\}.    
                                                   \nonumber
\end{eqnarray}

\begin{eqnarray}
 S^{+}(z_{n}, z'_{n}) 
 &=& {1\over 8\sqrt 2\pi l^2}
  \Biggl\{
          {{(r^2 -r^2_{-})^{1\over 2}
            ({r'}^2 -r^2_{-})^{1\over 2}}
           \over 
            {r^2_{+} - r^2_{-}} }
          \cosh F_{-}(\Delta t, \Delta \phi_{n}) 
                                                    \nonumber  \\
 &&     + {{(r^2_{+} -r^2)^{1\over 2}
            (r^2_{+} -{r'}^2)^{1\over 2}}
           \over 
            {r^2_{+} - r^2_{-}} }
          \cosh F_{+}(\Delta t, \Delta \phi_{n})
        - 1
  \Biggr\}^{-{3\over 2}}
                                                    \nonumber    \\
 &&\times 
  {{(r^2 -r^2_{-})^{1\over 4}
    ({r'}^2 -r^2_{-})^{1\over 4}}
   \over 
   {(r^2_{+} - r^2_{-})^{1\over 2}}}
  \exp{{1\over 2}
  \Bigl[ F_{-}(t, \phi_{n}) + F_{-}(t', \phi'_{n}) \Bigr]}
                                                    \nonumber    \\
 &&\times
  \Bigl[ - \gamma^0(z_0 - {z'}_0) 
         + \gamma^1(z_1 - {z'}_1) 
         + \gamma^2(z_2 - {z'}_2) 
  \Bigr].       
                                                    \nonumber
\end{eqnarray}
(3) For $z,z'\in $ Region (III); ${\rm 0} < r,r' \le  r_{-}$

\begin{eqnarray}
 (z_0 - {z'}_0) 
 &=& \Bigg\{ 
            \Bigl( 
                  {{r^2_{-} - r^2}\over {r^2_{+} - r^2}}
            \Bigr)^{1\over 2}
            \sinh F_{-}(t, \phi_{n}) e^{[-F_{+}(t, \phi_{n})]}
          - \Bigl( 
                  {{r^2_{-} - {r'}^2}\over {r^2_{+} - {r'}^2} }
            \Bigr)^{1\over 2}
             \sinh F_{-}(t', \phi'_{n}) e^{[-F_{+}(t', \phi'_{n})]}
     \Bigg\},     
                                                   \nonumber \\
 (z_1 - {z'}_1) 
 &=& \Bigg\{ 
            \Bigl( 
                  {{r^2_{-} - r^2}\over {r^2_{+} - r^2} }
            \Bigr)^{1\over 2}
            \cosh F_{-}(t, \phi_{n}) e^{[-F_{+}(t, \phi_{n})]}
          - \Bigl( 
                  {{r^2_{-} - {r'}^2}\over {r^2_{+} - {r'}^2} }
            \Bigr)^{1\over 2}
             \cosh F_{-}(t', \phi'_{n}) e^{[-F_{+}(t', \phi'_{n})]}
     \Bigg\},                                
                                                   \nonumber \\ 
 (z_2 - {z'}_2) 
 &=& \Bigg\{ 
            \Bigl( 
                  {{r^2_{+} - r^2_{-}}\over {r^2_{+} - r^2}}
            \Bigr)^{1\over 2}
            e^{[-F_{+}(t, \phi_{n})]} 
          - \Bigl( 
                  {{r^2_{+} - r^2_{-}}\over {r^2_{+} - {r'}^2}}
            \Bigr)^{1\over 2}
            e^{[-F_{+}(t', \phi'_{n})]}
     \Bigg\}.    
                                                   \nonumber
\end{eqnarray}

\begin{eqnarray}
 S^{+}(z_{n}, z'_{n}) 
 &=& {1\over 8\sqrt 2\pi l^2}
  \Biggl\{
          {{(r^2_{+} -r^2)^{1\over 2}
            (r^2_{+} -{r'}^2)^{1\over 2}}
           \over 
            {r^2_{+} - r^2_{-}} }
           \cosh F_{+}(\Delta t, \Delta \phi_{n})
                                                    \nonumber    \\   
 &&     - {{(r^2_{-} -r^2)^{1\over 2}
            (r^2_{-} -{r'}^2)^{1\over 2}}
           \over 
            {r^2_{+} - r^2_{-}} }
            \cosh F_{-}(\Delta t, \Delta \phi_{n})
        - 1
  \Biggr\}^{-{3\over 2}}
                                                    \nonumber    \\
 && \times {{(r^2_{+} -r^2)^{1\over 4}
    (r^2_{+} -{r'}^2)^{1\over 4}}
   \over 
   {(r^2_{+} - r^2_{-})^{1\over 2}}}
  \exp{{1\over 2}
  \Bigl[ F_{+}(t, \phi_{n}) + F_{+}(t', \phi'_{n}) \Bigr]}
                                                    \nonumber    \\
 &&\times
  \Bigl[ - \gamma^0(z_0 - {z'}_0) 
         + \gamma^1(z_1 - {z'}_1) 
         + \gamma^2(z_2 - {z'}_2) 
  \Bigr].       
                                                    \nonumber
\end{eqnarray}
(4) For $ z\in $ Region(I); $(r \ge r_{+})\ \ \ $ 
        $ z'\in $ Region (II); $ (r_{-} <  r' \le r_{+})$  

\begin{eqnarray}
 (z_0 - {z'}_0) 
 &=& \Bigg\{ 
            \Bigl( 
                  {{r^2 - r^2_{+}}\over {r^2 - r^2_{-}}}
            \Bigr)^{1\over 2}
            \sinh F_{+}(t, \phi_{n}) e^{[-F_{-}(t, \phi_{n})]}
          - \Bigl( 
                  {{r^2_{+} - {r'}^2}\over {{r'}^2 - r^2_{-}} }
            \Bigr)^{1\over 2}
             \cosh F_{+}(t', \phi'_{n}) e^{[-F_{-}(t', \phi'_{n})]}
     \Bigg\},     
                                                   \nonumber \\
 (z_1 - {z'}_1) 
 &=& \Bigg\{ 
            \Bigl( 
                  {{r^2 - r^2_{+}}\over {r^2 - r^2_{-}} }
            \Bigr)^{1\over 2}
            \cosh F_{+}(t, \phi_{n}) e^{[-F_{-}(t, \phi_{n})]}
          - \Bigl({{r^2_{+} - {r'}^2}\over {{r'}^2 - r^2_{-}} }
            \Bigr)^{1\over 2}
            \sinh F_{+}(t', \phi'_{n}) e^{[-F_{-}(t', \phi'_{n})]}
     \Bigg\},                                
                                                   \nonumber \\ 
 (z_2 - {z'}_2) 
 &=& \Bigg\{ 
            \Bigl( 
                  {{r^2_{+} - r^2_{-}}\over {r^2 - r^2_{-}}}
            \Bigr)^{1\over 2}
            e^{[-F_{-}(t, \phi_{n})]}
          - \Bigl( 
                  {{r^2_{+} - r^2_{-}}\over {{r'}^2 - r^2_{-}}}
            \Bigr)^{1\over 2}
            e^{[-F_{-}(t', \phi'_{n})]}
     \Bigg\}.    
                                                   \nonumber
\end{eqnarray}

\begin{eqnarray}
 S^{+}(z_{n}, z'_{n}) 
 &=& {1\over 8\sqrt 2\pi l^2}
  \Biggl\{
          {{(r^2 -r^2_{-})^{1\over 2}
            ({r'}^2 -r^2_{-})^{1\over 2}}
           \over 
            {r^2_{+} - r^2_{-}} }
          \cosh F_{-}(\Delta t, \Delta \phi_{n})
                                                    \nonumber    \\   
 &&     + {{(r^2 -r^2_{+})^{1\over 2}
            (r^2_{+} -{r'}^2)^{1\over 2}}
           \over 
            {r^2_{+} - r^2_{-}} }
            \sinh F_{+}(\Delta t, \Delta \phi_{n})
        - 1
  \Biggr\}^{-{3\over 2}}
                                                    \nonumber    \\
 &&\times 
  {{(r^2 -r^2_{-})^{1\over 4}
    ({r'}^2 -r^2_{-})^{1\over 4}}
   \over 
   {(r^2_{+} - r^2_{-})^{1\over 2}}}
  \exp{{1\over 2}
  \Bigl[ F_{-}(t, \phi_{n}) + F_{-}(t', \phi'_{n}) \Bigr]}
                                                    \nonumber    \\
 &&\times
  \Bigl[ - \gamma^0(z_0 - {z'}_0) 
         + \gamma^1(z_1 - {z'}_1) 
         + \gamma^2(z_2 - {z'}_2) 
  \Bigr].       
                                                    \nonumber
\end{eqnarray}
(5) For  $ z\in $ Region(II); $(r_{-} < r \le r_{+})\ \ \ $ 
         $ z'\in $ Region (III); $ ({\rm 0} < r' \le r_{-} )$ 

\begin{eqnarray}
 (z_0 - {z'}_0) 
 &=& \Bigg\{ 
            \Bigl( 
                  {{r^2_{+} - r^2}\over {r^2 - r_{-}^2}}
            \Bigr)^{1\over 2}
            \cosh F_{+}(t, \phi_{n}) e^{[-F_{-}(t, \phi_{n})]}
         -  \Bigl( 
                  {{r^2_{-} - {r'}^2}\over {r^2_{+} - {r'}^2} }
            \Bigr)^{1\over 2}
            \sinh F_{-}(t', \phi'_{n}) e^{[-F_{+}(t', \phi'_{n})]}
     \Bigg\},     
                                                   \nonumber \\
 (z_1 - {z'}_1) 
 &=& \Bigg\{ 
            \Bigl( 
                  {{r^2_{+} - r^2}\over {r^2 - r_{-}^2} }
            \Bigr)^{1\over 2}
             \sinh F_{+}(t, \phi_{n}) e^{[-F_{-}(t, \phi_{n})]}
         - \Bigl( 
                  {{r^2_{-} - {r'}^2}\over {r^2_{+} - {r'}^2} }
            \Bigr)^{1\over 2}
            \cosh F_{-}(t', \phi'_{n}) e^{[-F_{+}(t', \phi'_{n})]}
     \Bigg\},                                
                                                   \nonumber \\ 
 (z_2 - {z'}_2) 
 &=& \Bigg\{ 
            \Bigl( 
                  {{r^2_{+} - r^2_{-}}\over {r^2 - r_{-}^2}}
            \Bigr)^{1\over 2}
             e^{[-F_{-}(t, \phi_{n})]}
         - \Bigl( 
                  {{r^2_{+} - r^2_{-}}\over {r^2_{+} - {r'}^2}}
            \Bigr)^{1\over 2}
             e^{[-F_{+}(t', \phi'_{n})]}
     \Bigg\}.    
                                                   \nonumber
\end{eqnarray}

\begin{eqnarray}
 S^{+}(z_{n}, z'_{n}) 
 &=& {1\over 8\sqrt 2\pi l^2}
  \Biggl\{
          {{(r^2 -r^2_{-})^{1\over 2}
            (r^2_{+} -{r'}^2)^{1\over 2}}
           \over 
            {r^2_{+} - r^2_{-}} }
           \cosh [(a_{-}t-a_{+}t') - l(a_{+}\phi_{n} - a_{-}\phi'_{n})]
                                                    \nonumber    \\   
 &&     - {{(r^2_{+} -r^2)^{1\over 2}
            (r^2_{-} -{r'}^2)^{1\over 2}}
           \over 
            {r^2_{+} - r^2_{-}} }
           \sinh [(a_{+}t-a_{-}t') - l(a_{-}\phi_{n} - a_{+}\phi'_{n})]
        - 1
  \Biggr\}^{-{3\over 2}}
                                                    \nonumber    \\
 && \times {{(r^2 -r^2_{-})^{1\over 4}
    (r^2_{+} -{r'}^2)^{1\over 4}}
   \over 
   {(r^2_{+} - r^2_{-})^{1\over 2}}}
  \exp{{1\over 2}
  \Bigl[ F_{-}(t, \phi_{n}) + F_{+}(t', \phi'_{n}) \Bigr]}
                                                    \nonumber    \\
 &&\times
  \Bigl[ - \gamma^0(z_0 - {z'}_0) 
         + \gamma^1(z_1 - {z'}_1) 
         + \gamma^2(z_2 - {z'}_2) 
  \Bigr].       
                                                    \nonumber
\end{eqnarray}
where again we used the short-hand notation, 
$F_{\pm}(t, \phi_{n}) \equiv [a_{\pm}t - la_{\mp}\phi_{n}]$,
$F_{\pm}(\Delta t, \Delta \phi_{n}) \equiv 
[a_{\pm}(t-t') - la_{\mp}(\phi-\phi' + 2\pi n)]$
and where in (2+1)-dimension, Dirac $\gamma$-matrices  
$ \gamma^{a} = (\gamma^0, \gamma^1, \gamma^2)$ 
obeying
\begin{eqnarray}
 (\gamma^0)^{\dagger} &=& \gamma^0, 
 (\gamma^i)^{\dagger} = -\gamma^i,              \nonumber \\
 (\gamma^0)^2 &=& I, (\gamma^i)^2 = -I,         \nonumber \\
 \{ \gamma^a , \gamma^b \} &=& -2\eta^{ab}      \nonumber
\end{eqnarray}
are given, for example,  in standard representation, by

\begin{eqnarray}
 \gamma^0 &=& \sigma_3 = \left(\matrix{1&0\cr 0&-1\cr}\right),
                                                  \nonumber \\
 \gamma^1 &=& i\sigma_1 = i\left(\matrix{0&1\cr 1&0\cr}\right) 
                        =  \left(\matrix{0&i\cr i&0\cr}\right),
                                                  \nonumber \\ 
 \gamma^2 &=& i\sigma_2 = i\left(\matrix{0&-i\cr  i&0\cr}\right) 
                        =  \left(\matrix{0&1 \cr -1&0\cr}\right).
                                                  \nonumber 
\end{eqnarray}
This completes the explicit evaluation of Green's functions for 
the scalar and spinor field in the background of AdS$_3$ black hole
spacetime. \\
Perhaps it would be appropriate to mention the issue of boundary
conditions at infinity on any Green's function for fields propagating
on general AdS$_3$ spacetimes. As is the case with all AdS$_3$
spacetimes, the AdS$_3$ black hole spacetimes of BTZ [3] we are dealing
with is {\it not} globally hyperbolic. And this global non-hyperbolicity
invites some delicacy in constructing the two-point Green's functions
which is under the consideration. To be more concrete, the spatial
infinity $i^{0}$ of this AdS$_3$ black hole spacetime is ``timelike''.
This has been demonstrated in the Carter-Penrose conformal diagram
provided by BTZ [3]. And physically, it means that information can leak in
or out through the spatial infinity in a finite coordinate time. In
order to deal with this fact, boundary conditions may be imposed on
the two-point Green's functions. As has been pointed out by Avis,
Isham and Storey [12], it is still possible to define a quantization scheme
on globally non-hyperbolic spacetimes like the present AdS$_3$ black hole
spacetime without using boundary conditions, which is usually referred to
as ``transparent'' boundary conditions. In view of this, the construction
of two-point Green's functions for conformally-coupled scalar and spinor
field performed in the present work can be thought of as corresponding
to this option.
The two-point Green's functions for quantum fields in AdS$_3$ black hole
spacetime have been provided in the literature [16]. But only the Green's
functions for conformally-coupled scalar field with the choice of Neumann
or Dirichlet boundary conditions are given. Thus the Green's functions for
conformally-coupled spinor field given in the present work can be regarded
as a new contribution. 

\vspace*{0.5cm}

\noindent

\begin{center}
{\large\bf References}
\end{center}

\begin{description}

\item {[1]} C. W. Misner, K. S. Thorne, and J. A. Wheeler, {\it Gravitation}
            (San Francisco : Freeman, 1973).

\item {[2]} N. D. Birrel, and P. C. W. Davies, 
            {\it Quantum fields in curved space} 
            (Cambridge Univ. Press, 1982).

\item {[3]} M. Banados, C. Teitelboim, and J. Zanelli, Phys. Rev. 
            Lett. {\bf 69}, 1849 (1992) ; M. Banados, M. Henneaux,
            C. Teitelboim, and J. Zanelli, Phys. Rev. {\bf D48},
            1506 (1993).
            
\item {[4]} S. Carlip, ``The (2+1)-Dimensional Black Hole'', gr-qc/9506079.

\item {[5]} S. Hyun, G. H. Lee and J. H. Yee, Phys. Lett. {\bf B322}, 182
            (1994) ; S. Hyun, Y. S. Song and J. H. Yee, Phys. Rev. {\bf D51},
            1787 (1995).

\item {[6]} R. Penrose, Riv. Nuovo Cimento {\bf 1}, 252 (1969).

\item {[7]} C. W. Misner, Phys. Rev. Lett. {\bf 28}, 994 (1972) ;
            W. H. Press and S. A. Teukolsky, Nature {\bf 238}, 211 (1972) ;
            Ya. B. Zeldovich, J. Exp. Theor. Phys. {\bf 62}, 2076 (1972).

\item {[8]} S. Chandrasekhar, and S. Detweiler, Proc. R. Soc. Lond.
            {\bf A352}, 325 (1976) ; S. Chandrasekhar, Proc. R. Soc. Lond.
            {\bf A348}, 39 (1976) ;  Proc. R. Soc. Lond. {\bf A350}, 165
            (1976) ; {\it The Mathmatical Theory of Black Holes} (Oxford,
            Oxford University Press, 1983).

\item {[9]} W. Unruh, Phys. Rev. Lett. {\bf 31}, 1265 (1973) ; R. Guven,
            Phys. Rev. {\bf D16}, 1706 (1977).

\item {[10]} R. M. Wald, {\it General Relativity} (Univ. of Chicago Press,
             Chicago, 1984).

\item {[11]} Eisenhardt, {\it Riemannian Geometry} (Princeton Univ. Press, 1949).

\item {[12]} S. J. Avis, C. J. Isham, and D. Storey, Phys. Rev. {\bf D18},
             3565 (1978).

\item {[13]} S. W. Hawking, Commun. Math. Phys. {\bf 43}, 199 (1975).

\item {[14]} W. G. Unruh, Phys. Rev. {\bf D14}, 870 (1976).

\item {[15]} B. S. DeWitt, in {\it General Relativity, an Einstein Centenary
             Survey}, eds. S. W. Hawking and
             W. Israel (Cambridge, Cambridge University Press, 1979).

\item {[16]} A. R. Steif, Phys. Rev. {\bf D49}, R585 (1994) ; 
             G. Lifschytz, and M. Ortiz, Phys. Rev. {\bf D49}, 1929 (1994) ;
             I. Ichnose, and Y. Satoh, Nucl. Phys. {\bf B447}, 340 (1995). 

\item {[17]} D. Birmingham, I. Sachs, and S. Sen, Phys. Lett. {\bf B413},
             281 (1997) ; E. Teo, hep-th/9805014.

\item {[18]} H. W. Lee, N. J. Kim, and Y. S. Myung, hep-th/9803080.

\item {[19]} A. Dasgupta, hep-th/9808086.

\end{description}

\end{document}